\documentclass[copyright,creativecommons,sharealike]{eptcs}
\providecommand{\event}{DSL 2011} 

\usepackage[utf8]{inputenc}
\usepackage{breakurl}             
\usepackage{relsize}
\usepackage{xspace}
\usepackage{graphicx}
\usepackage{pstricks}
\usepackage{epsfig}
\usepackage{alltt}
\usepackage{pifont}
\usepackage{amssymb}
\usepackage{fancyvrb}

\newcommand{\gccmelt}{\textsf{\textsc{gcc}\textsuperscript{melt}}\xspace}
\newcommand{\MELT}{\textsf{MELT}\xspace}

\newcommand{\prodname}[1]{\textsf{\relsize{-0.5}#1}}
\newcommand{\melt}{\prodname{MELT}\xspace}
\newcommand{\gcc}{\prodname{GCC}\xspace}
\newcommand{\ggc}{\prodname{Gg-c}\xspace}
\newcommand{\kloc}{\textsc{kloc}\xspace}
\newcommand{\Mloc}{\textsc{Mloc}\xspace}

\newenvironment{myalltt}{\relsize{-1}\begin{alltt}}{\end{alltt}}
\newcommand{\mytexttt}[1]{{\relsize{-0.5}\texttt{#1}}}

\newcommand{\mymelt}[1]{{\relsize{-0.4}\texttt{\textsc{#1}}}}
\newcommand{\myboldcode}[1]{{\relsize{-0.5}\texttt{\textbf{#1}}}}
\newcommand{\meltcodec}[1]{{\relsize{-0.5}{{\textbf{\texttt{#1}}}}}}

\newcommand{\petitun}[1]{{\relsize{-1}{#1}}}
\newcommand{\petitdemi}[1]{{\relsize{-0.5}{#1}}}
\newcommand{\quasicode}[1]{\(\lceil\)\textrm{#1}\(\rfloor\)}

\urldef\basilenetemail\url{basile@starynkevitch.net}
\urldef\basileceaemail\url{basile.starynkevitch@cea.fr}

\title{\emph{MELT}\\ a Translated Domain Specific Language \\
Embedded in the ~ \emph{GCC} ~ Compiler}
\author{Basile \textsc{Starynkevitch}
\institute{CEA, LIST\\
\relsize{-1}{Software Safety Laboratory, boîte courrier 94, 91191 \textsc{Gif/Yvette Cedex}, France}}
\email{\quad basile@starynkevitch.net \quad\qquad basile.starynkevitch@cea.fr}
}
\def\titlerunning{\emph{MELT} - a translated DSL inside the \emph{GCC} compiler}
\def\authorrunning{B.Starynkevitch}
\begin{document}
\maketitle

\begin{abstract}
The GCC free compiler is a very large software, compiling source in several
languages for many targets on various systems. It can be extended by
plugins, which may take advantage of its power to provide extra
specific functionality (warnings, optimizations, source refactoring or
navigation) by processing various GCC internal representations
(Gimple, Tree, ...). Writing plugins in C is a complex and
time-consuming task, but customizing GCC by using an existing
scripting language inside is impractical. We describe \emph{MELT}, a
specific Lisp-like DSL which fits well into existing GCC technology
and offers high-level features (functional, object or reflexive
programming, pattern matching). MELT is translated to C fitted for GCC
internals and provides various features to facilitate this. This work
shows that even huge, legacy, software can be a posteriori extended by
specifically tailored and translated high-level DSLs.
\end{abstract}

\section{Introduction}
\label{sec:intro}

\gcc\footnote{Gnu Compiler Collection (\texttt{gcc 4.6.0} released on
  march 25\textsuperscript{th} 2011) on \url{gcc.gnu.org}} is an
industrial-strength free compiler for many source languages (C, C++,
Ada, Objective C, Fortran, Go, ...), targetting about 30 different
machine architectures, and supported on many operating systems. Its
source code size is huge (4.296\Mloc \footnote{4.296 Millions Lines Of
  source Code, measured with David Wheeler's
  \prodname{SLOCCount}. Most other tools give bigger code measures,
  e.g., \texttt{ohcount} gives 8.370\Mloc of source, with 5.477\Mloc of
  code and 1.689\Mloc of comments.} for \gcc 4.6.0), heterogenous, and
still increasing by 6\% annually\,\footnote{\gcc 4.4.1, released July
  22\textsuperscript{th}, 2009, was 3.884\Mloc, so a 0.412\Mloc =
  10.6\% increase in 1.67 years}. It has no single main architect and hundreds of (mostly
full-time) contributors, who follow strict
social rules~\footnote{Every submitted code patch should be accepted
  by a code reviewer who cannot be the author of the patch, but there
  is no project leader or head architect, like Linus Torvalds is for
  the Linux kernel. So \gcc has not a clean, well-designed,
  architecture.}.

\subsection{The powerful GCC legacy}
\label{subsec:gcc-legacy}

The several \gcc \cite{gcc-internals} front-ends (parsing C, C++, Go
\ldots source) produce common internal \emph{AST} {(abstract syntax
  tree)} representations called \emph{Tree} and \emph{Generic}. These
are later transformed into middle-end internal representations, the
\emph{Gimple} statements - through a transformation called
\emph{gimplification}. The bulk of the compiler is its
\emph{middle-end} which operates repeatedly on these \emph{Gimple}
representations\footnote{The \gcc middle-end does not depend upon the
  source language or the target processor (except with parameters
  giving \texttt{sizeof(int)} etc.).}. It contains nearly 200 passes
moulding these (in different forms). Finally, back-ends (specific to
the target) work on \emph{Register Transfer Language} (RTL)
representations and emit assembly code. Besides that, many other data
structures exist within \gcc (and a lot of global variables). Most of
the compiler code and optimizations work by various transformations on
middle-end internal representations. \gcc source code is mostly written
in \emph{C} (with a few parts in C++, or Ada), but it also has several internal
\emph{C} code generators. \gcc does not use parser generators
(like \petitun{\texttt{flex}, \texttt{bison}}, etc).

It should be stressed that \emph{most} internal \gcc
\emph{representations} are constantly \emph{evolving}, and \emph{there
  is no stability}\footnote{This is nearly a dogma of its community,
  to discourage proprietary software abuse of \gcc.} of the internal
\gcc API\footnote{\gcc has no well defined and documented Application
  Programming Interface for compiler extensions; its API is just a big
  set of header files, so is a bit messy for outsiders.}. This makes
the embedding of existing scripting languages (like Python, Ocaml,
\ldots) impractical (\S
\ref{subsec:embedding-existing-language}). Since \texttt{gcc 4.5} it
is possible to enhance \gcc through external \emph{plugins}.

External \emph{plugins} can enhance or modify the behavior of the \gcc
compiler through a defined interface, practically provided by a set of \emph{C}
file headers, and made of functions, many \emph{C} macros, and
coding conventions. Plugins are loaded as \texttt{dlopen}-ed dynamic
shared objects at \texttt{gcc} run time. They can profit from all the variety
and power of the many internal representations and processing of
\gcc. Plugins enhance \gcc by inserting new passes and/or by
responding to a set of plugin events (like
\mytexttt{PLUGIN\_FINISH\_TYPE} when a type has been parsed,
\mytexttt{PLUGIN\_PRAGMAS} to register new pragmas, \ldots).

\gcc plugins can add specific warnings (e.g., to a library), specific
optimizations (e.g., transform \mytexttt{fprintf(stdout,\textrm{...})}
$\rightarrow$ \mytexttt{printf(\textrm{...})} in user code with
\mytexttt{\#include <stdio.h>}), compute software metrics, help on
source code navigation or code refactoring, etc. \gcc extensions or
plugins enable using and extending \gcc for non code-generation
activities like static analysis
\cite{glek-2008-gcc-grepsed,CousotCousot92-2,monate-08-slicing,venet-array-check-2004},
threats detection (like in
\prodname{Two}\cite{abstr-int-threat-2001}, \prodname{Coverity\texttrademark}~\footnote{See
  \href{http://www.coverity.com}{\texttt{www.coverity.com}}}, or
\prodname{Astrée}\cite{CousotEtAl-ASTREE-ESOP05,CousotEtAl06-ASIAN}),
code refactoring, coding rules
validation\cite{marpons-codingrules-2008}, etc. They could provide any
processing taking advantage of the many facilities already existing
inside \gcc. However, since coding \gcc plugins in \emph{C} is not
easy, a higher-level DSL could help. Because \gcc plugins are usually
specific to a narrow user community, shortening their development time
(through a higher-level language) makes sense.

\begin{figure}[h]
\begin{myalltt}
\textit{/* A node in a gimple_seq_d.  */}
struct \textbf{GTY}((chain_next ("%h.next"), chain_prev ("%h.prev"))) gimple_seq_node_d \{
  gimple stmt;
  struct gimple_seq_node_d *prev;
  struct gimple_seq_node_d *next; \};
\end{myalltt} 
\hspace{5em} \petitun{(code from \texttt{gcc/gimple.h} in \gcc)}
\caption{example of \texttt{GTY} annotation for \ggc}
\label{fig:example-GTY}
\end{figure}

\medskip Since compilers handle many complex (perhaps circular) data
structures for their internal representations, explicitly managing
memory is cumbersome during compilation. So the \gcc community has
added a crude \emph{garbage collector} \cite{jones1996garbage} \ggc ~ {(\gcc Garbage
  Collector)}: many \emph{C} \texttt{struct}-ures in \gcc code are
annotated with \texttt{\textbf{GTY}} (figure \ref{fig:example-GTY}) to
be handled by \ggc; passes can allocate them, and a
precise \footnote{\ggc is a \emph{precise} G-C knowing each pointer to
  handle; using Boehm's conservative garbage collector with ambigous
  roots inside \gcc has been considered and rejected on performance
  grounds.} mark and sweep garbage collection may be triggered by the
pass manager \emph{only between passes}. \ggc does not know about
\emph{local} pointers, so garbage collected data is live and kept only
if it is (indirectly) reachable from known global or static
\texttt{GTY}-annotated variables (data reachable only from local
variables would be lost). Data internal to a \gcc pass is usually
manually allocated and freed. \texttt{GTY} annotations on types and
variables inside \gcc source are processed by \texttt{gengtype}, a
specialized generator (producing \emph{C} code for \ggc allocation and
marking routines and roots registration). There are more than 1800
\texttt{GTY}-ed types known by \ggc, such as: \mytexttt{gimple}
(pointing to the representation of a \emph{Gimple} statement),
\mytexttt{tree} (pointer to a structure representing a \emph{Tree}),
\mytexttt{basic\_block} (pointing the representation of a basic block
of Gimple-s), \mytexttt{edge} (pointing to the data representing an
edge between basic blocks in the control flow graph), etc.  Sadly, not
all \gcc data is handled by \ggc; a lot of data is still manually
micro-managed. We call \textbf{\relsize{+1}{\emph{stuff}}} all the
\gcc internal data, either garbage-collected and
\texttt{GTY}-annotated like \texttt{gimple}, \texttt{tree}, \ldots, or
outside the heap like raw \texttt{long} numbers, or even manually
allocated like \texttt{struct opt\_pass} (data describing \gcc
optimization passes).

\medskip

\gcc is a big legacy system, so its API is large and quite
heterogenous in style. It is not only made of data declarations and
functions operating on them, but also contains various \emph{C}
macros. In particular, iterations inside internal representations may
be provided by various styles of constructs:
\begin{enumerate}

\item Iterator abstract types like (to iterate on every \texttt{stmt},
  a gimple inside a given basic block \texttt{bb}) \begin{small}\begin{alltt}
  \textbf{for} (gimple_stmt_iterator gsi = gsi_start_bb (bb);
       !gsi_end_p (gsi); gsi_next (&gsi)) \{
    gimple stmt = gsi_stmt (gsi); \textit{/* handle \textrm{stmt} ...*/} \}
\end{alltt}\end{small}
\item Iterative \texttt{for}-like macros, e.g., (to iterate for each basic block \texttt{bb} inside the current function \texttt{cfun})\begin{small}\begin{alltt}
basic_block bb;   FOR_EACH_BB (bb) \{ \textit{/* process \textrm{bb} */} \} 
\end{alltt}\end{small}

\item More rarely, passing a callback to an iterating ``higher-order'' \emph{C}
  function, e.g., (to iterate inside every index tree from \texttt{ref}
  and call \texttt{idx\_infer\_loop\_bounds} on that index tree)
\begin{small}\begin{alltt}for_each_index (&ref, idx_infer_loop_bounds, &data);
\end{alltt}\end{small} with a static function \mytexttt{bool
idx\_infer\_loop\_bounds (tree base, tree *idx, void *dta)} 
called on every index tree \texttt{base}.
\end{enumerate}

\subsection{Embedding an existing scripting language is impractical}
\label{subsec:embedding-existing-language}

Interfacing \gcc to an existing language implementation like Ocaml,
Python, Guile, Lua, Ruby or some other scripting language is not
realistic~\footnote{The author spent more than a month of work trying in vain
  to plug Ocaml into \gcc!}, because of an \emph{impedance
  mismatch}: \begin{enumerate}
\item \label{enum:lang-gc} Most scripting languages are garbage collected, and \emph{mixing
  several garbage collectors is difficult and error-prone}, in
  particular when both \ggc and scripting language heaps are intermixed.
\item \label{enum:big-api} The \gcc API is very big, ill-defined, heterogenous, and
  evolving significantly. So manually coding the glue code between
  \gcc and a general-purpose scripting language is a big burden, and
  would be obsoleted by a new \gcc version
  when achieved.
\item \label{enum:weird-api} The \gcc API is not only made of \emph{C} functions, but also
  of macros which are not easy to call from a scripting language.
\item \label{enum:lowlevel-api} Part of the \gcc API is very low-level (e.g., field accessors),
  and would be invoked very often, so may become a performance
  bottleneck if used through a glue routine.
\item  \label{enum:various-data} \gcc handles various internal data (notably using hundreds of
  global variables), some through \texttt{GTY}-ed \ggc collected pointers
  (like \mytexttt{gimple\_seq}, \mytexttt{edge}, \ldots), others
  with manually allocated data (e.g., \mytexttt{omp\_region} for OpenMP
  parallel region information) or with numbers mapping some opaque
  information (e.g., \mytexttt{location\_t} are integers encoding
  source file locations). \gcc data has widely different types, usage
  conventions, or liveness.
\item \label{enum:no-root-type} There is no single root type (e.g., a
  root class like \mytexttt{GObject} \footnote{See
    \url{http://developer.gnome.org/gobject/}} in \prodname{Gtk})
  which would facilitate gluing \gcc into a dynamically typed language
  interpreter (à la Python, Guile, or Ruby).
\item  \label{enum:no-static-type} Statically typing \gcc data into a strongly typed language with
  type inference like Ocaml or Haskell is impractical, since it would
  require the formalization of a type theory compatible with all the
  actual \gcc code.
\item \label{enum:matching} Easily filtering complex nested data
  structures is very useful inside compilers, so most \gcc
  extensions need to \emph{pattern-match} on existing \gcc \emph{stuff} (notably
  on \emph{Gimple} or \emph{Tree}-s).
\end{enumerate}

The \melt (originally meaning \emph{``Middle End Lisp Translator''})
Domain Specific Language has been developped to increase, as any
high-level DSL does, the programmer's productivity.  \melt has its
specific generational copying garbage collector above \ggc to address
point \ref{enum:lang-gc}. Oddity of the \gcc API (points
\ref{enum:big-api}, \ref{enum:weird-api}, \ref{enum:lowlevel-api}) is
handled by generating well fit \emph{C} code, and by providing
mechanisms to ease that \emph{C} source code generation. Items
\ref{enum:lowlevel-api}, \ref{enum:various-data},
\ref{enum:no-root-type}, \ref{enum:no-static-type} are tackled by
mixing \melt dynamically typed values with raw \gcc
\emph{stuff}. \melt has a powerful pattern matching ability to handle
last point \ref{enum:matching}, because scripting languages don't
offer extensible or embeddable pattern matching (on data structures
internal to the embedding application).

\melt is being used for various \gcc extensions (work in progress):
\begin{itemize}
\item simple warning and optimization like
  \mytexttt{fprintf(stdout,\textrm{...})} detection and transformation
  (handling it on \emph{Gimple} representation is preferable to simple
  textual replacement, because it cooperates with the compiler
  inlining transformation);
\item Jérémie Salvucci has coded  a \emph{Gimple} $\rightarrow$ \emph{C}  transformer (to feed some other tool);
\item Pierre Vittet is coding various domain-specific warnings (e.g., detection of untested calls to \texttt{fopen});
\item the author is developing an extension to generate \emph{OpenCL} code
  from some \emph{Gimple}, to transport some highly parallel regular
  (e.g., matrix) code to GPUs;
\end{itemize}

\subsection{MELT = a DSL translated to code friendly with GCC internals}

The legacy constraints given by \gcc on additional (e.g., plugins')
code suggest that a DSL for extending it could be implemented by
generating \emph{C} code suitable for \gcc internals, and by providing
language constructs translatable into \emph{C} code conforming to \gcc
coding style and conventions. Other attempts to embed a scripting
language into \gcc (Javascript \cite{glek-2008-gcc-grepsed} for coding
rules in \prodname{Firefox}, Haskell for enhancing C++ template
meta-programming \cite{collingbourne-2009-haskellgcc}, or
Python\footnote{See David Malcom's \gcc Python plugin announced in
  \url{http://gcc.gnu.org/ml/gcc/2011-06/msg00293.html}} ) have
restricted themselves to a tiny part of the \gcc API;
Volanschi \cite{volanschi-mygcc-2006} describes a modified \gcc compiler with
specialized matching rules.

Therefore, \textbf{the reasonable way to provide a higher-level domain
  specific language for \gcc extensions is to dynamically generate
  suitable \emph{C} code} adapted to \gcc's style and legacy and similar in
form to existing hand-coded \emph{C} routines inside \gcc. This is the
driving idea of our \melt domain specific language and plugin
implementation
\cite{starynkevitch-melt,starynkevitch-2007-gcc,starynkevitch-2009-grow}.
By generating suitable \emph{C} code for \gcc internals, \melt fits well into
existing \gcc technology. This is in sharp contrast with the
\prodname{Emacs} editor or the
\prodname{\mbox{C\hspace{0.15ex}{-}\hspace{0.15ex}{-}}}~
compiler \cite{ramsey-cminusminus-2000} whose architecture was
designed and centered on an embedded interpreter (E-Lisp for
\prodname{Emacs}, Lua\textsuperscript{ocaml} for
\prodname{\mbox{C\hspace{0.15ex}{-}\hspace{0.15ex}{-}}}).

\melt is a Lisp-looking DSL designed to work on \gcc internals. It handles
both dynamically typed \melt values and raw \gcc \emph{stuff} (like
\mytexttt{gimple}, \mytexttt{tree}, \mytexttt{edge} and many
others). It supports applicative, object and reflective programming
styles. It offers powerful pattern matching facilities to work on \gcc
internal representations, essential inside a compiler. It is
translated into \emph{C} code and offer linguistic devices to deal nicely
with \gcc legacy code.


\section{Using \MELT and its runtime.}
\label{sec:using-runtime}

\subsection{\MELT usage and organization overview}
From the user's perspective, the \gcc compiler enabled with \melt
(\gccmelt) can be run with a command as: \mytexttt{gcc -fplugin=melt
  -fplugin-arg-melt-mode=opengpu -O -c foo.c}. This instructs
\mytexttt{gcc} (the \mytexttt{gcc-4.6} packaged in Debian) to run the
compiler proper \mytexttt{cc1}, asks it to load the \mytexttt{melt.so}
plugin which provides the \melt specific runtime infrastructure, and
passes to that plugin the argument \mytexttt{mode=opengpu} while
\mytexttt{cc1} compiles the user's \mytexttt{foo.c}.  The
\mytexttt{melt.so} plugin initializes the \melt runtime, hence itself
\mytexttt{dlopen}-s \melt modules like \mytexttt{warmelt*.so} \&
\mytexttt{xtramelt*.so}. These modules initialize \melt data,
e.g., classes, instances, closures, and handlers. The \melt handler associated to the
\mytexttt{opengpu} mode registers a new \gcc pass (available in
\mytexttt{xtramelt-opengpu.melt}) which is
executed by the \gcc pass manager when compiling the file
\mytexttt{foo.c}.  This \mytexttt{opengpu} pass uses Graphite
\cite{TRIFUNOVIC:2010:INRIA-00551516:1} to find optimization
opportunities in loops and should \footnote{In April 2011, the
  \mytexttt{opengpu} pass, coded in \melt, is still incomplete in
  \melt 0.7 svn rev.173182.}  generate OpenCL code to run these on
GPUs, transforming the Gimple to call that generated OpenCL code.  The
\mytexttt{melt.so} plugin is mostly hand-coded in C (in our
\mytexttt{melt-runtime.[hc]} files - 15\kloc, which
\mytexttt{\#include} generated files).  The \melt modules
\mytexttt{warmelt*.so} \& \mytexttt{xtramelt*.so} \footnote{The
  module names \mytexttt{warmelt*.so} \& \mytexttt{xtramelt*.so}
  are somehow indirectly hard-coded in \mytexttt{melt-runtime.c} but
  could be overloaded by many explicit \mytexttt{-fplugin-arg-melt-*}
  options.}  are coded in \melt (as source files
\mytexttt{warmelt*.melt}, \ldots, \mytexttt{xtramelt*.melt} which have
been translated by \melt into generated C files \mytexttt{warmelt*.c}
\& \mytexttt{xtramelt-*.c}, themselves compiled into modules
\mytexttt{warmelt*.so} \ldots).

\sloppypar
The \melt translator (able to generate \mytexttt{*.c} from
\mytexttt{*.melt}) is \emph{bootstrapped} so that it exercises most of its
features and its runtime : the translator's source code is coded in
\melt , precisely the \mytexttt{melt/warmelt*.melt} files (39\kloc), and
the \melt \emph{source} repository also contains
the \emph{generated} files \mytexttt{melt/generated/warmelt*.c} (769\kloc). Other \melt
files, like \mytexttt{melt/xtramelt*.melt} (6\kloc{}) don't need to have their generated translation kept. The \melt
translator\footnote{The translation from file \mytexttt{ana-simple.melt}
  to \mytexttt{ana-simple.c} is done by invoking \mytexttt{gcc
    -fplugin=melt -fplugin-arg-melt-mode=translatefile
    -fplugin-arg-melt-arg=ana-simple.melt} \ldots on an
  \emph{empty} C file \mytexttt{empty.c}, only to have
  \mytexttt{cc1} launched by \mytexttt{gcc}!}  is \emph {not} a \gcc
front-end (since it produces C code for the host system, not
\emph{Generic} or \emph{Gimple} internal representations suited for
the target machine); and it is even able to dynamically generate,
during an \gccmelt compiler invocation, some temporary \mytexttt{*.c}
code, run \mytexttt{make} to compile that into a temporary
\mytexttt{*.so}, and load (i.e. \mytexttt{dlopen}) and execute that - all
this in a single \mytexttt{gcc} user invocation; this can be useful for
sophisticated static analysis \cite{starynkevitch-2007-gcc} specialized using
partial evaluation techniques within the
analyzer, or just to ``run'' a \melt file.

The \melt translator works in several steps: the reader builds
s-expressions in \melt heap. Macro-expansion translates them into a
\melt AST. Normalization introduces necessary temporaries and builds a
normal form. Generation makes a representation very close to \emph{C}
code. At last that representation is emitted to output generated
\emph{C} code. There is no optimization done by the \melt translator
(except for compilation of pattern matching, see \S \ref{sub:implementing-patterns}).

Translation from \melt code to C code is fast: on a x86-64
\textsf{GNU/Linux} desktop system\footnote{An Intel Q9550 @ 2.83GHz,
  8Gb RAM, fast 10KRPM Sata 150Gb disk, \textsf{Debian/Sid/AMD64}.},
the 6.5\kloc~ \mytexttt{warmelt-normal.melt} file is translated into
five \mytexttt{warmelt-normal*.c} files with a total of 239\kloc in just
one second (wall time). But 32 seconds are needed to build the
\mytexttt{warmelt-normal.so} module (with \mytexttt{make}\footnote{So
  it helps to run that in parallel using \mytexttt{make -j}; the 32
  seconds timing is a sequential single-job \texttt{make}.}  running
\texttt{gcc -O1 -fPIC}) from these generated C files. So most of the
time is spent in compiling the generated C code, not in generating
it. In contrast to several DSLs persisting their
closures\footnote{Ocaml bytecode contains both code and
  data; GNU \{ Emacs, CLisp, Smalltalk \} persist their entire heap
  image. But \melt has no persistent data files, to avoid serializing
  \gcc's stuff (ie \gcc's native data).}  by serializing a mixture of data and code, \melt
starts with an empty heap, so \melt modules' initialization routines
are mostly long and sequential C code initializing the \melt heap.

\subsection{MELT runtime infrastructure}

The \melt runtime \mytexttt{melt-runtime.c} is built above the \gcc
infrastructure, notably \ggc. However, \ggc is not a sufficient
garbage collector for \melt values, like closures, lists, tuples,
objects, ... As in most applicative or functional languages, \melt
code tends to allocate a lot of temporary values (which often die
quickly). So garbage collection (G-C) of \melt values may happen
often, and does need to happen even inside \gcc passes written in
\melt, not only between passes. These values are handled by our
\emph{generational copying} \melt G-C, triggered by the \melt
allocator when its birth region is full, and backed up by the existing
\ggc (so the old generation of \melt G-C is the \ggc
heap). Generational copying GCs \cite{jones1996garbage} handle quickly
dead young temporary values by discarding them at once after having
copied each live young value out of the birth region, but require a
scan of all local variables, need to forward pointers to moved values,
a write barrier, and \emph{normalization} (like the
\emph{administrative normal form} in
\cite{Flanagan:2004:ECC:989393.989443}) of explicit intermediate
values inside calls\footnote{That is, $f(g(x),y)$ should be normalized
  as $\tau = g(x); f(\tau,y)$ with $\tau$ being a fresh
  temporary.}. This is awkward in hand-written C code but easy to
generate. Minor \melt G-Cs are triggered before each call to
\texttt{gcc\_collect} (i.e. to the full \ggc) to ensure that all live
young \melt values have migrated to the old \ggc heap. Compatibility
between our \melt GC and \ggc is thus achieved. An array of more than
a hundred \emph{predefined} values contains the only ``global'' \melt
values (which are global roots for both the \melt GC and \ggc).

\melt call frames are aggregated as local \texttt{struct}-ures,
containing local \melt values, the currently called \melt closure, and
local stuff (like raw \texttt{tree} pointers, etc.). Values inside these
call frames are known to the \melt garbage collector, which scans them
and possibly moves them. Expliciting these call frames facilitates
\emph{introspective runtime reflection}
\cite{pitrat-oneself-1995,pitrat-96-reflective,pitrat-beings-2009} at the \melt level;
this might be useful for some future sophisticated analysis, e.g., in
abstract interpretation \cite{CousotCousot92-2,CousotCousot04-WCC} of
recursive functions, as a widening strategy. Concretely, local \melt
values (and stuff) are aggregated in \melt call frames (represented as
generated \emph{C} local \mytexttt{struct}-ures) organized in a
single-linked list. This also enables the display of the \melt
backtrace stack on errors.

\begin{figure}
\begin{relsize}{-1}
\begin{myalltt}
melt_ptr_t meltgc_new_int (meltobject_ptr_t discr_p, long num) \{
  MELT_ENTERFRAME (2, NULL);
#define newintv meltfram__.mcfr_varptr[0]
#define discrv  meltfram__.mcfr_varptr[1]
  discrv = (void *) discr_p;
  if (melt_magic_discr ((melt_ptr_t) (discrv)) != MELTOBMAG_OBJECT)
    goto end;
  if (((meltobject_ptr_t)discrv)->obj_num != MELTOBMAG_INT)
    goto end;
  newintv = meltgc_allocate (sizeof (struct meltint_st), 0);
  ((struct meltint_st*)newintv)->discr = (meltobject_ptr_t)discrv;
  ((struct meltint_st*)newintv)->val = num;
end:
  MELT_EXITFRAME ();
  return (melt_ptr_t) newintv;
\}
\end{myalltt}
\end{relsize}

\smallskip
\caption{\label{fig:newboxint} \melt runtime function boxing an integer}
\end{figure}

\sloppypar
The figure \ref{fig:newboxint} gives an example of hand-written code
following \melt conventions (a function \mytexttt{meltgc\_new\_int} boxing an integer into a value
of given dicriminant and number to be boxed). It uses the
\mytexttt{MELT\_ENTERFRAME} macro\footnote{The \emph{Ocaml} runtime has similar macros.}, which is expanded by the \emph{C}
preprocessor into the code in figure \ref{fig:enterframe}, which
declares and initialize the \melt call frame
\mytexttt{meltfram\_\_}. The \mytexttt{MELT\_EXITFRAME ()} macro
occurrence is expanded into \mytexttt{melt\_topframe = (struct
  melt\_callframe\_st *) (meltfram\_\_.mcfr\_prev);} to pop the
current \melt frame. \melt provides a \gcc pass checking some of \melt
coding conventions in the hand-written part of the \melt runtime.
\begin{figure}
\begin{relsize}{-1}
\begin{myalltt}
  struct \{
    int mcfr_nbvar;                          \textit{/* number of \melt local values*/}
    const char *mcfr_flocs;                  \textit{/* location string for debugging*/}
    struct meltclosure_st *mcfr_clos;        \textit{/* current closure*/}
    struct melt_callframe_st *mcfr_prev;     \textit{/* link to previous \melt frame */}
    void *mcfr_varptr[2];                    \textit{/* local \melt values */}
  \} meltfram__;        \textit{/* \melt current call frame */}
  static char locbuf_1591[84];        \textit{/* location string */}
  if (!locbuf_1591[0])
     snprintf (locbuf_1591, sizeof (locbuf_1591) - 1, "%s:%d",  basename ("gcc/melt-runtime.c"), (int) 1591);
  memset (&meltfram__, 0, sizeof (meltfram__));
  meltfram__.mcfr_nbvar = (2);
  meltfram__.mcfr_flocs = locbuf_1591;
  meltfram__.mcfr_prev = (struct melt_callframe_st *) melt_topframe;
  meltfram__.mcfr_clos = (((void *) 0));
  melt_topframe = ((struct melt_callframe_st *) &meltfram__);
\end{myalltt}
\end{relsize}
\smallskip
\caption{\label{fig:enterframe} C preprocessor expansion of \mytexttt{MELT\_ENTERFRAME(2, NULL)} at line 1591}

\end{figure}

The \melt runtime depends deeply upon \ggc, but does not depend much
on the details of \gcc's main data structures like e.g., \mytexttt{tree}
or \mytexttt{gimple} or \mytexttt{loop} : our
\mytexttt{melt-runtime.c} can usually be recompiled without changes
when \gcc's file \mytexttt{gimple.h} or \mytexttt{tree.h} changes, or
when passes are changed or added in \gcc's core. The \melt translator
files \mytexttt{warmelt*.melt} (and the generated
\mytexttt{warmelt*.c} files) don't depend really on \gcc data
structures like \mytexttt{gimple}. As a case in point, \emph{the major
  ``gimple to tuple'' transition} \footnote{In the old days of \gcc
  version 4.3 the
  \emph{Gimple} representation was physically implemented in
  \mytexttt{tree}-s and the C data structure \mytexttt{gimple} did not
  exist yet; at that time, \emph{Gimple} was sharing the same physical
  structures as \emph{Tree}s and \emph{Generic} [so \emph{Gimple} was mostly a conventional
    restriction on \emph{Tree}s] - that is using many linked lists. The 4.4
  release added the \mytexttt{gimple} structure to represent them,
  using arrays, not lists, for sibling nodes; this improved
  significantly \gcc's performance but required patching many
  files.}  in \texttt{gcc-4.4}, which impacted a lot of \gcc files, \emph{was
  smoothly handled} within the \melt translator.

The \melt files which are actually processing \gcc internal
representations (like our \mytexttt{xtramelt-*.melt} or user \melt code),
that is \melt code implementing new \gcc passes, have to change only when the
\gcc API changes - exactly like other \gcc passes. Often, since the
change is compatible with existing code, these \melt files don't have
to be changed at all (but should be recompiled into modules).

\melt handles two kinds of \textit{\textbf{thing}}s: the first-class \melt
\emph{value}s (allocated and managed in \melt's GC-ed heap) and other
\emph{stuff}, which are any other \gcc data managed in C (either generated
or hand-written C code within \gccmelt). Informally, $\textit{Things} = \textit{Values} \cup \textit{Stuff}$. So raw \mytexttt{long}-s,
\mytexttt{edge}-s or \mytexttt{tree}-s are \emph{stuff}, and appear
exactly in \melt memory like C-coded \gcc passes handle them (without
extra boxing). Variables and [sub-]expressions in \melt code, hence
locals in \melt call frames, can be \emph{thing}s of either kind (values or \emph{stuff}).

Since \ggc requires each pointer to be of a \mytexttt{gengtype}-
known type, values are really different from \emph{stuff}. There is
unfortunately \emph{no way to implement full polymorphism} in \melt:
we cannot have \melt tuples containing a mix of raw
\mytexttt{tree}-s and \melt objects (even if both are \ggc managed
pointers). This \ggc limitation has deep consequences in the \melt
language (\emph{stuff}, i.e. \gcc native data, sadly cannot be first-class
\melt values!). 

Some parts of the \melt runtime are generated (by a special \melt
mode). Various \melt values' and \emph{stuff} implementation are described
by \melt instances. So adding extra types of values, or interfacing
additional \gcc \emph{stuff} to \melt{}, is fairly simple, but requires
a complete re-building of \melt. Their \mytexttt{GTY((\textrm{...}))}
\mytexttt{struct}-ure declarations in \emph{C} are generated. Lower
parts of the \melt runtime (allocating, forwarding, scanning routines
- see chapters 6 \& 7 of \cite{jones1996garbage} - for the copying
\melt G-C, hash-tables implementation, \ldots) are also
generated. This generated \emph{C} code is kept in the source
repository.

Notice that the distinction between first-class \melt \emph{value}s and plain
\emph{stuff} is essential in \melt, and is required by current \gcc practices
(notably its \ggc collector). Therefore, the \melt language itself
needs to denote them separately and explicitly, and the \melt runtime
(and generated code) handles them differently. In that respect, \melt
is \emph{not} like Lisp, Scheme, Guile, Lua and Python. However, \melt
coders should usually prefer handling values (the ``first class
citizens''), not raw \emph{stuff}.

\subsection{\MELT debugging aids}

When generating non-trivial \emph{C} code, it is important to lower
the risk of crashing the generated code \footnote{However, it is still
  possible to make some \melt code crash, for instance by adding bugs
  in the \emph{C} form of our code chunks \S
  \ref{subsub:code-chunk}. In practice, \melt code crashes very
  rarely; most often it fails by breaking some assertions.}. This is
achieved by systematically clearing all data (both values and raw
\emph{stuff}) to avoid uninitialized pointers (and \melt G-C also requires
that), and by carefully coding low-level operations (primitives \S
\ref{subsub:primitives}, c-matchers \S \ref{sub:c-matchers}, code
chunks \S \ref{subsub:code-chunk}) with tests against null pointers.

The generated \emph{C} code produced by the \melt translator contains
many \mytexttt{\#line} directives (suitably wrapped with
\texttt{\#ifdef}).  In the rare cases when the \texttt{gdb} debugger
needs to be used on \melt code (e.g., to deal with crashes or infinite
loops), it will refer correctly to the originating \melt source file
location. These positions are also written into \melt call frames, to ease backtracing on error.

\melt uses debug printing and assertions quite extensively. If enabled
by the \mytexttt{-fplugin-arg-melt-debug} program argument to
\mytexttt{gcc}, a lot of debug printing happens : each use of the
\mytexttt{debug\_msg} operation displays the current \melt source
location, a message, and a value \footnote{Values are printed for
  debug use with \melt message passing through the
  \mymelt{dbg\_output} \& \mymelt{dbg\_outputagain} selectors.}. For
debugging \emph{stuff} data, primitives \mytexttt{debugtree},
\mytexttt{debuggimple}, etc. are available.  Assertions are provided
by \mytexttt{assert\_msg} which takes a message and a condition to
check. When the check fails, the entire \melt call stack is printed
(with positions referring to \mytexttt{*.melt} source files).

When variadic functions will be available in \melt, their first use
will support polymorphic debug printing. A \mytexttt{debug} ``macro''
would be expanded into calls to a \mytexttt{debug\_at} variading
function, which would get the source location value as its first
argument, and the values or \emph{stuff} to be debug-printed as secondary
variadic arguments.

An older version of \melt could be used with an external probe, which
was a graphical program interacting with \texttt{cc1} through
asynchronous textual protocols. This approach required a quite
invasive patch of \gcc's code itself. The current \gcc pass manager
and plugin machinery now provides enough hooks, and future versions of
\melt might communicate asynchronously with a central monitor (to be
developed).


\section{The \MELT language and its pecularities}
\label{sec:language}

Some familiarity with a Lisp-like language (like Emacs Lisp, Scheme,
Common Lisp, etc.) is welcome to understand this section. Acquaintance
with a dynamically typed scripting language like Python, Guile or Ruby
could also help. See the web site
\href{http://gcc-melt.org/}{\texttt{gcc-melt.org}} for more material
(notably tutorials) on \melt.

\melt has a Lisp-like syntax because it was (at its very beginning)
implemented with an initial ``external'' \melt to C translator
prototyped in Common Lisp. Since then, a lot of newer features have
been progressively added (using an older version of \melt to bootstrap
its current version). The \prodname{Emacs Lisp} language (in the
\prodname{Emacs} editor), \prodname{Guile} (the \prodname{Gnu} implementation of \emph{Scheme}), and \emph{machine
  description} files in \gcc back-end are successful examples of other Lisp
dialects within \prodname{Gnu} software. Finally, existing editing
modes\footnote{\prodname{Emacs} mode for \prodname{Lisp} is nearly
  enough for editing, highlighting and indenting \melt code.} for
\prodname{Lisp} are sufficient for \melt.

An alternative infix syntax (code-named \prodname{Milt}) for \melt is
in the works; the idea is to have an infix parser, coded in \melt, for
future \texttt{*.milt} files, which is parsed into \melt internal
s-expressions (i.e. into the same instances of \mymelt{class\_sexpr}
as the \melt Lisp-like reader does): symbols starting with \texttt{+}
or \texttt{-} are parsed as infix operators (like Ocaml does) with
additive precedences, those starting with \texttt{*} or \texttt{/}
have multiplicative precedence, etc.

\melt shares with existing Lisp languages many syntactic and lexical
conventions for comments, indentation, symbols (which may be non alpha-numerical), case-insensitivity, and a lot
of syntax (like \texttt{if}, \texttt{let}, \texttt{letrec},
\texttt{defun}, \texttt{cond} \ldots). As in all Lisp dialects, everything is parenthesized like \mytexttt{(
  \textit{operator} \textit{operands \textrm{...}\,})} so parenthesis
are highly significant. The quote, back-quote, comma and question mark characters have special
significance, so \texttt{\large 'a} is parsed exactly as
\mytexttt{(quote a)}, \texttt{\large ?b} as \mytexttt{(question b)}
etc. Like in Common Lisp, words prefixed with a colon like \texttt{:long}
are considered as ``keywords'' and are not subject to
evaluation. Symbols and keywords exist both in source files and in the
running \melt heap.

\medskip

\subsection{\MELT macro-strings}
\label{sub:macro-strings}
Since ``mixing'' C code chunks  (\S \ref{subsub:code-chunk}) inside \melt code is very important,
simple meta-programming is implemented by a 
lexical trick \footnote{Inspired by handling of
  \texttt{\$} in strings or ``here-documents'' by shells, Perl, Ruby, ...}:
\emph{macro-strings} are strings prefixed with
     {\texttt{\textbf{\#\{}}} and suffixed with
     {\texttt{\textbf{\}\#}}} and are parsed specially; these prefix
     and suffix strings have been chosen because they usually don't
     appear in \emph{C} code. Within a macro-string,
     backslash does not escape characters, but \texttt{\$}
     and sometimes \texttt{\#} are scanned specially, to parse symbols
     inside macro-strings. \\ For example, \melt reads the
     macro-string \hspace{1em} {
       \verb!#{/*$P#A*/printf("a=%ld\n", $A);}#! } \hspace{2em}
     exactly as a list
     \verb!("/*"  p  "A*/printf(\"a=%ld\\n\", "  a  ");")!  \hspace{1em}
     of 5 elements whose 1\textsuperscript{st}, 3\textsuperscript{rd}
     and 5\textsuperscript{th} elements are strings\footnote{The first
       string has the two characters \textit{\tt /*} and the last has
       the two characters \textit{\tt );}} and 2\textsuperscript{nd}
     and 4\textsuperscript{th} elements are symbols \texttt{p} and
     \texttt{a}.  This is useful when one wants to mix C code inside
     \melt code; some macro-strings are several dozens of lines long,
     but don't need any extra escapes (as would be required by using
     plain strings).

Another example of \emph{macro-string} is given in the following
``hello-world'' (complete) \melt program:

\begin{myalltt}\textit{;; file helloworld.melt}
(\textbf{code\_chunk} helloworldchunk 
          #\{int i=0; /* our \$HELLOWORLDCHUNK */ 
            \$HELLOWORLDCHUNK#_label: printf("hello world from MELT\textbackslash{n}");
            if (i++ < 3) goto \$HELLOWORLDCHUNK#_label; \}#)
\end{myalltt}

The macro-string spans on 3 lines, and contains some \emph{C}
code with the \texttt{helloworldchunk} \melt symbol. The above \texttt{helloworld.melt}
file (of 4 lines) is translated into a \texttt{helloworld.c} file (of 389
lines\footnote{With 260 lines of code, including 111 preprocessor
  directives, mostly \texttt{\#line}, and 129 comment or blank lines, and all the code doing
  ``initialization''.} in \emph{C}). It uses the \texttt{code\_chunk}
construct explained in \S \ref{subsub:code-chunk} below (to emit
translated \emph{C} code).

\subsection{\MELT values and stuff}
\label{subsec:values}

Every \melt \emph{value} has a \emph{discriminant} (at the start of the
memory zone containing that value). As an exception, nil \footnote{As
  in Common Lisp or Emacs Lisp (or C itself), but not as in Scheme,
  \melt nil value is considered as false, and every non-nil value is true.},
represented by the C null pointer has conventionally a specific
discriminant \mymelt{discr\_null\_reciever}.  The discriminant of a value is
used by the \melt runtime, by \ggc and in \melt code to separate
them. \melt values can be boxed \emph{stuff} (e.g., boxed \mytexttt{long} or
boxed \mytexttt{tree}), closures, lists, pairs, tuples, boxed strings,
\ldots, and \melt \emph{objects}.  Several \emph{predefined objects},
e.g., \mymelt{class\_class}, \mymelt{discr\_null\_receiver} \ldots,
are required by the \melt runtime. The hierarchy of
discriminants is rooted at
\mymelt{discr\_any\_receiver} \footnote{\mymelt{discr\_any\_receiver}
  is rarely used, e.g., to install catch-all method
  handlers.}. Discriminants are objects (of
\mymelt{class\_discriminant}). Core classes and discriminants are
predefined as \melt values (known by both \ggc and \melt G-C).

Each \melt object has its class as its discriminant. Classes are
themselves objects and are organized in a single-inheritance hierarchy
rooted at \mymelt{class\_root} (whose parent discriminant is
\mymelt{discr\_any\_reciever}). Objects are represented in C as
exactly a structure with its class (i.e. discriminant)
\mytexttt{obj\_class}, its unsigned hash-code \mytexttt{obj\_hash}
(initialized once and for all), an unsigned ``magic'' short number
\mytexttt{obj\_num}, the unsigned short number of fields
\mytexttt{obj\_len}, and the \mytexttt{obj\_vartab[obj\_len]} array
of fields, which are \melt values. The \mytexttt{obj\_num} in objects
can be set \emph{at most once} to a \emph{non-zero} unsigned short,
and may be used as a \emph{tag}: \melt and \ggc discriminate quickly
a value's data-type (for marking, scanning and other purposes) through
the \mytexttt{obj\_num} of their discriminant. So, safely testing in \emph{C}
if a value \mytexttt{p} is a \melt closure is as fast as \mytexttt{p
  != NULL \&\& p->discr->obj\_num == MELTOBMAG\_CLOSURE}.

\melt field descriptors and method selectors are objects. Every \melt
\emph{value} (object or not, even nil) can be sent a message, since its
discriminant (i.e., its class, if it is an object) has a method map (a
hash table associating selectors to method bodies) and a parent
discriminant (or super-class). Message passing in \melt is similar to
those in \prodname{Smalltalk} and \prodname{Ruby}. Method bodies can
be dynamically installed with \mytexttt{(install\_method \textrm{\it
    discriminant selector function})} and removed at any time in any
discriminant or class. Method invocations use the method hash-maps
(similar to methods' dictionnaries in \prodname{Smalltalk}) to find the
actual method to run.

The \melt reader produces mostly objects and sometimes other values:
S-expressions are parsed as instances of \mymelt{class\_sexpr}
(containing the expression's source location and the list of its
components); symbols (like \mymelt{==} or \myboldcode{let} or
\mytexttt{x}) as instances of \mymelt{class\_symbol}; keywords like
\myboldcode{:long} or \myboldcode{:else} as instances of
\mymelt{class\_keyword}; numbers like \mytexttt{-1} as values of
\mymelt{discr\_integer} etc.

Each \emph{stuff} (that is, non-value \emph{thing}s like \mytexttt{long} or
\mytexttt{tree} \ldots) have its boxed value counterpart, so boxed
gimple-s are values containing, in addition of their discriminant
(like \mymelt{discr\_gimple}), a raw \mytexttt{gimple} pointer.

In \melt expressions, literal integers like \mytexttt{23} or strings
like \mytexttt{"hello\textbackslash{n}"} refer to raw
\myboldcode{:long} or \myboldcode{:cstring} \emph{stuff} \footnote{All
  \myboldcode{:cstring} are \mytexttt{(const char*)} C-strings in the
  text segment of the executable, so they are not
  \mytexttt{malloc}-ed.}, not constant values. To be considered
as \melt values they need to be quoted, so (contrarily to other Lisps) in
\melt ~ \mytexttt{2 $\not\equiv$ '2} : the plain \mytexttt{2} denotes
a raw stuff of c-type \myboldcode{:long} so is not a value, but the
quoted expression \mytexttt{'2} denotes the boxed integer 2 constant
value of \mymelt{discr\_constant\_integer} so they are not equivalent!
As in Lisp, a quoted symbol like \mytexttt{'j} denotes a constant value
(of \mymelt{class\_symbol}).

\medskip
To associate \emph{thing}s (either \melt objects or \gcc \emph{stuff}, all of the
same type) to \melt values, hash-maps are extensively used: so
homogenous hash tables keyed by objects, raw strings, or raw \emph{stuff}
like \mytexttt{tree}-s or \mytexttt{gimple}-s \ldots are values (of
discriminant \mymelt{discr\_map\_objects} \ldots,
\mymelt{discr\_map\_trees}). While hash-maps are more costly than
direct fields in structures to associate some data to these
structures, they have the important benefit of avoiding disturbing
existing data structures of \gcc. And even C plugins of \gcc cannot
add for their own convenience extra fields into the carefully tuned
\mytexttt{tree} or \mytexttt{gimple} structures of \gcc's
\mytexttt{tree.h} or \mytexttt{gimple.h}.

Aggregate \melt values include not only objects, hash-tables and
pairs, but also tuples (a value containing a fixed number of immutable
component values), closures, lists, \ldots Lists know their first and
last pairs. Aggregate values of the same kind may have various
discriminants. For instance, within a \melt class (which is itself a
\melt object of \mymelt{class\_class}) a field gives the tuple of all
super-classes starting with \mymelt{class\_root}. That tuple has
\mymelt{discr\_class\_sequence} as discriminant, while most other
tuples have \mymelt{discr\_multiple} as discriminant.

\emph{Decaying values} may help algorithms using memoization; they
contain a value reference and a counter, decremented at each major
garbage collection. When the counter reaches 0, the reference is
cleared to nil.

\medskip

Adding a new important \gcc C type like
\mytexttt{gimple} \footnote{This kind of radical addition don't happen
  often in the \gcc community because it usually impacts a lot of \gcc
  files.}  for some new stuff is fairly simple: add (in \melt code) a
new predefined C-type descriptor (like \mymelt{ctype\_gimple}
referring to keyword \myboldcode{:gimple}) and additional
discriminants, and regenerate all of \melt. C-type descriptors
(e.g., \mymelt{ctype\_edge}) and value type descriptors (like
\mymelt{valdesc\_list}) contains dozen[s] of fields (names or body
chunk of generated \emph{C} routines) used when generating the runtime
support routines.

The \myboldcode{:void} keyword (and so \mymelt{ctype\_void}) is used
for side-effecting code without results. C-type keywords (like
\myboldcode{:void}, \myboldcode{:long}, \myboldcode{:tree},
\myboldcode{:value}, \myboldcode{:gimple}, \myboldcode{:gimple\_seq},
etc.) qualify (in \melt source code) formal arguments, local
variables (bound by \myboldcode{let}, \ldots), etc.

\melt is typed for \emph{thing}s: e.g., the translator complains if the
\mytexttt{+i} primitive addition operator (expecting two raw
\myboldcode{:long} \emph{stuff} and giving a \myboldcode{:long} result) is
given a value or a \myboldcode{:tree} argument. Furthermore, \myboldcode{let}
bindings can be explicitly typed (by default they bind a
value). Within values, typing is dynamic; for instance, a value is
checked at runtime to be a closure before being applied. When applying
a \melt closure to arguments, the first argument, if any, needs to be
a value (it would be the receiver if the closure is a method for
message passing)\footnote{The somehow arbitrary requirement of having
  the first argument of every \melt function be a value speeds up calls
  to functions with one single value argument, and permits using
  closures as methods without checks: sending a
  message to a raw \emph{stuff} like e.g., a \texttt{tree} won't work.}, others can be
\emph{thing}s, i.e. values or \emph{stuff}. In \melt applications, the types of
secondary arguments and secondary results are described by constant
byte strings, and the secondary arguments or results are passed (in
generated \emph{C} code) as an array of unions. The generated \melt
function prologue (in \emph{C}) checks that the formal and actual type of
secondary arguments are the same (otherwise, argument passing stops,
and all following actual arguments are cleared).

All \melt \emph{thing}s (value or \emph{stuff}), in particular local variables (or
mismatched formals), are initially cleared (usually by zeroing the
whole \melt call frame in the \emph{C} prologue of each generated
routine). So \melt values are initially \texttt{()} (i.e., nil in \melt
syntax), a \mytexttt{:tree} \emph{stuff} is initially the null tree
(i.e. \texttt{(tree)0} in \emph{C} syntax), a \mytexttt{:long} \emph{stuff} is
initially \texttt{0L}, a \mytexttt{:cstring} \emph{stuff} is initialized to
\texttt{(const char*)0}. Notice that cleared \emph{stuff} is considered as
false in conditional context.

\medskip
Functions written in \melt (with \myboldcode{defun} for named
functions or \myboldcode{lambda} for anonymous ones) always return a
value as their primary result (which may be ignored by the caller, and
defaults to nil). The first formal argument (if any) and the primary
result of \melt functions should be values (so nested function calls
deal mainly with values). Secondary arguments and results can be any
\emph{thing}s (each one is either a value or some \emph{stuff}). The
\mytexttt{(\textbf{multicall} \textrm{...})} syntax binds primary and
secondary results like Common Lisp's \myboldcode{multiple-value-bind}.

\subsection{Syntax overview}

The following constructs should be familiar (except the last one,
\texttt{match}, for pattern matching) since they look like in other
Lisps. Notice that our \mytexttt{let} is always sequential\footnote{So
  the \meltcodec{let} of \melt is like the \texttt{\textbf{let*}} of
  Scheme!}. Formals in abstractions \footnote{Notice that
  \myboldcode{lambda} abstractions are constructive expressions and
  may appear in \myboldcode{letrec} or \myboldcode{let} bindings.} are
restricted to start with a formal value; this speeds up the common
case of functions with a single value argument, and facilitates installation of
any function as method (without checking that the formal reciever is
indeed a value).

List of formal arguments (in \myboldcode{lambda}, \myboldcode{defun}
etc.) contains either symbols (which are names of formals bound by
e.g., the \myboldcode{lambda}) like \mytexttt{x} or \mytexttt{discr},
or c-type keywords like \myboldcode{:value} or \myboldcode{:long} or
\myboldcode{:gimple} \ldots. A c-type keyword qualify all
successing formals up to the next c-type keywords, and the default
c-type is \myboldcode{:value}. For example, the formal arguments list
\mytexttt{(x y \myboldcode{:long} n k \myboldcode{:gimple} g
  \myboldcode{:value} v)} have 6 formals : \mytexttt{x y v} are
\melt values, \mytexttt{n k} are raw long \emph{stuff}, \mytexttt{g} is a raw
gimple \emph{stuff}.

Local bindings (in \myboldcode{let} or \myboldcode{letrec}) has an
optional c-type annotation, then the newly bound symbol, then the
sub-expression bounding it. So \mytexttt{(\textbf{:long} x 2)} locally
binds (in the body of the enclosing \myboldcode{let}) the symbol
\mytexttt{x} to the raw long \emph{stuff} \texttt{2}, and in the \texttt{let}
body \texttt{x} is a raw long variable.

\medskip
Patterns and pattern matching are explained in \S\ref{sec:pattern}.

\bigskip
\begin{small}
\begin{tabular}{l|p{0.25\textwidth}|p{0.46\textwidth}}
\multicolumn{3}{c}{\textbf{expressions} where $n \ge 0$ and $p \ge 0$} \\
\hline
application & \texttt{($\phi$ $\alpha_1$ \textrm{...} $\alpha_n$)} &
apply function \petitun{(or primitive)} $\phi$ to arguments $\alpha_i$ \\
\hline
assignment & \texttt{(\myboldcode{setq} $\nu$ $\epsilon$)} & set local
  variable $\nu$ to $\epsilon$ \\
\hline
message passing & \texttt{($\sigma$ $\rho$ $\alpha_1$ \textrm{...}
  $\alpha_n$)} & send selector $\sigma$ to reciever $\rho$ with
arguments $\alpha_i$ \\
\hline
let expression &  \texttt{(\myboldcode{let}
  ($\beta_1$\textrm{...}$\beta_n$)
  $\epsilon_1$\textrm{...}$\epsilon_p$ $\epsilon'$)} & with local
\textbf{sequential} bindings $\beta_i$ evaluate
side-effecting sub-expressions $\epsilon_j$ and give result of $\epsilon'$\\
\hline
sequence &  \texttt{(\myboldcode{progn} 
  $\epsilon_1$\textrm{...}$\epsilon_n$ $\epsilon'$)} & evaluate
$\epsilon_i$ \petitun{(for their side effects)} and at last $\epsilon'$, giving its
result \petitun{(like the operator \textbf{\texttt{,}} in C)} \\
\hline
abstraction & \texttt{(\myboldcode{lambda} $\phi$
    $\epsilon_1$\textrm{...}$\epsilon_n$ $\epsilon'$)} & anonymous function with
  formals $\phi$ and side-effecting expressions $\epsilon_i$, return
  result of $\epsilon'$ \\
\hline
\textbf{pattern matching} & \texttt{(\myboldcode{match} $\epsilon$ $\chi_1$ \textrm{...} $\chi_n$)} & match
result of $\epsilon$ against match clauses $\chi_i$, giving result of
last expression of matched clause. \\
\end{tabular}
\end{small}
\bigskip

Conditional expressions alter control flow as usual. However,
conditions can be \emph{thing}s, e.g., the \texttt{0}
\myboldcode{:long} \emph{stuff} is false, other long \emph{stuff} are
true, a \texttt{gimple} \emph{stuff} is false iff it is the null gimple pointer,
etc. The ``else'' part $\epsilon$ of an \texttt{if} test is
optional. When missing, it is false, that is a cleared \emph{thing}.  Notice
that tested conditions and the result of a conditional expression can
be either values or raw stuff, but all the conditional sub-expressions
of a condition should have consistent types, otherwise the entire
expression has \texttt{:void} type.

\medskip
\begin{small}
\begin{tabular}{l|p{0.20\textwidth}|p{0.60\textwidth}}
\multicolumn{3}{c}{\textbf{conditional expressions} where $n \ge 0$ and $p \ge 0$} \\
\hline
test & \texttt{(\myboldcode{if} $\tau$ $\theta$
  $\epsilon$)} & if $\tau$ then $\theta$ else $\epsilon$ \petitun{(like
\textbf{\texttt{?:}} in C)} \\
\hline
conditional & \texttt{(\myboldcode{cond} $\kappa_1$ \textrm{...}
  $\kappa_n$)} & evaluate conditions $\kappa_i$ until one is satisfied \\
\hline
conjunction  & \texttt{(\myboldcode{and} $\kappa_1$ \textrm{...}
  $\kappa_n$ $\kappa'$)} & if $\kappa_1$ and then $\kappa_2$ \ldots
and then $\kappa_n$ is ``true'' (non nil or non zero) then $\kappa'$ otherwise
the cleared \emph{thing} of same type \\
\hline
disjunction & \texttt{(\myboldcode{or} $\delta_1$ \textrm{...}
  $\delta_n$)} & $\delta_1$ or else $\delta_2$ \ldots = the first of the $\delta_i$ which is ``true'' (non
nil, or non zero, ...) \\
\hline
\end{tabular}
\end{small}

\medskip
In a \myboldcode{cond} expression, every condition
$\kappa_i$ (except perhaps the last) is like \texttt{($\gamma_i$
  $\epsilon_{i,1}$ \textrm{...} $\epsilon_{i,p_i}$ $\epsilon'$)} with $p_i \ge
0$. The first such condition for which $\gamma_i$ is ``true'' gets its
sub-expressions $\epsilon_{i,j}$ evaluated sequentially for their
side-effects and gives the result of $\epsilon'$. The last condition can be
\texttt{(\myboldcode{:else} $\epsilon_1$ \textrm{...} $\epsilon_n$
  $\epsilon'$)}, is triggered if all previous conditions failed, and (with
the sub-expressions $\epsilon_i$ evaluated sequentially for their
side-effects) gives the result of $\epsilon'$

\bigskip

\melt has some more expressions.

\begin{small}
\begin{tabular}{l|p{0.35\textwidth}|p{0.35\textwidth}}
\multicolumn{3}{c}{\textbf{more expressions}} \\
\hline
loop & \texttt{(\myboldcode{forever} $\lambda$ $\alpha_1$ \textrm{...} $\alpha_n$)} &
loop indefinitely on the $\alpha_i$ which may exit \\
\hline
exit & \texttt{(\myboldcode{exit} $\lambda$ $\epsilon_1$ \textrm{...}
  $\epsilon_n$ $\epsilon'$)} & exit enclosing loop $\lambda$ after
side-effects of $\epsilon_i$ and result of $\epsilon'$ \\
\hline
return & \texttt{(\myboldcode{return} $\epsilon$ $\epsilon_1$ \textrm{...}
  $\epsilon_n$)} & return $\epsilon$ as the main result, and the
$\epsilon_i$ as secondary results \\
\hline
multiple call & \texttt{(\myboldcode{multicall} $\phi$ $\kappa$
  $\epsilon_1$\textrm{...}$\epsilon_n$ $\epsilon'$)} & locally bind formals $\phi$ to
\petitdemi{main and secondary} result[s] of \petitdemi{application or send} $\kappa$ and
evaluate the $\epsilon_i$ for side-effects and $\epsilon'$ for result \\
\hline
recursive let &  \texttt{(\myboldcode{letrec}
  ($\beta_1$\textrm{...}$\beta_n$)
  $\epsilon_1$\textrm{...}$\epsilon_p$)} & with [mutually-] recursive
\emph{constructive} bindings $\beta_i$ evaluate sub-expressions
$\epsilon_j$ \\
\hline
field access & \texttt{(\myboldcode{get\_field} :$\Phi$ $\epsilon$)} & if
$\epsilon$ gives an appropriate object retrieves its field $\Phi$,
otherwise nil \\
\hline
\textbf{unsafe} field access & \texttt{(\myboldcode{unsafe\_get\_field}
  :$\Phi$ $\epsilon$)} & unsafe access without check like the above operation\\
\hline
object update & \texttt{(\myboldcode{put\_fields} $\epsilon$  \textbf{:$\Phi_1$}
    $\epsilon_1$ \textrm{...} \textbf{:$\Phi_n$}
    $\epsilon_n$)} & safely update  (if appropriate) in the object given by
$\epsilon$ each field $\Phi_i$ with  $\epsilon_i$\\
\hline
unsafe object update & \texttt{(\myboldcode{unsafe\_put\_fields} $\epsilon$  \textbf{:$\Phi_1$}
    $\epsilon_1$ \textrm{...})} & unsafely update the object given by
$\epsilon$
\end{tabular}
\end{small}
\medskip

The unsafe field access \myboldcode{unsafe\_get\_field} is reserved to
expert \melt programmers, since it may crash. The safer variant test that the
expression $\epsilon$ evaluates\footnote{I.e. test if the value
  $\omega$ of $\epsilon$ is an object which is a direct or indirect
  instance of the class defining field $\Phi$, otherwise a nil value
  is given.} to a \melt object of appropriate class before accessing a
field $\Phi$ in it. Field updates with \myboldcode{put\_fields} are
safe \footnote{Update object $\omega$, value of $\epsilon$, only if it
  is an object which is a direct or indirect instance of the class
  defining each field $\Phi_i$}, with an unsafe but quicker variant \mytexttt{unsafe\_put\_fields} available for \melt experts.

\bigskip

Mutually recursive \myboldcode{letrec} bindings should have only
constructive expressions.

\begin{small}
\begin{tabular}{l|l|p{0.33\textwidth}}
\multicolumn{3}{c}{\textbf{constructive expressions}} \\
\hline
list & \texttt{(\myboldcode{list} $\alpha_1$ \textrm{...} $\alpha_n$)} &
make a list of $n$ values $\alpha_i$ \\
\hline
tuple & \texttt{(\myboldcode{tuple}  $\alpha_1$ \textrm{...} $\alpha_n$)} &
make a tuple of $n$ values  $\alpha_i$ \\
\hline
instance & \texttt{(\myboldcode{instance} $\kappa$ \textbf{:$\Phi_1$}
    $\epsilon_1$ \textrm{...} \textbf{:$\Phi_n$}
    $\epsilon_n$)} & make an instance of class $\kappa$ and $n$ fields
    $\Phi_i$ set to value $\epsilon_i$\\
\end{tabular}
\end{small}
\medskip

Of course \myboldcode{lambda} expressions are also constructive and
can appear inside \myboldcode{letrec}. Notice that since \melt is
translated into C, and because of runtime constraints, \melt recursion
is never handled tail-recursively so always consume stack space. This
also motivates iterative constructions (like \myboldcode{forever} and
our iterators).

\bigskip

Name defining expressions have a syntax starting with
\myboldcode{def}. Most of them (except \myboldcode{defun},
\myboldcode{defclass}, \myboldcode{definstance}) have no equivalent in
other languages, because they define bindings related to C code
generation. For the \melt translator, bindings have various kinds; each
binding kind is implemented as some subclass of
\mymelt{class\_any\_binding}.

Name exporting expressions are essentially directives for the module
system of \melt. Only exported names are visible outside a module. A
module initialization expects a parent environment and produces a
newer environment containing exported bindings.  Both name defining
and exporting expressions are supposed to appear only at the top-level
(and should not be nested inside other \melt expressions).
\smallskip

\begin{small}
\relsize{-0.8}
\begin{tabular}{l|p{0.35\textwidth}|p{0.40\textwidth}}
\multicolumn{3}{c}{\textbf{expressions defining names}} \\
for functions & \texttt{(\myboldcode{defun} $\nu$ $\phi$ $\epsilon_1$
  \textrm{...} $\epsilon_n$ $\epsilon'$)} & define function $\nu$ \petitdemi{with
formal arguments $\phi$ and body $\epsilon_1$
  \textrm{...} $\epsilon_n$ $\epsilon'$} \\
\hline
for classes &  \texttt{(\myboldcode{defclass} $\nu$ \myboldcode{:super} $\sigma$
  \myboldcode{:fields} ($\phi_1 \ldots \phi_n$) )} 
& define class $\nu$ of super-class $\sigma$ and own fields $\phi_i$ 
\\
\hline
for instances &  \texttt{(\myboldcode{definstance} $\iota$ $\kappa$
  :$f_1$ $\epsilon_1$ \textrm{...} :$f_n$ $\epsilon_n$)} 
& define an instance $\iota$ of class $\kappa$ with each field $f_i$
initialized to the value of $\epsilon_i$ 
\\
\hline
for selectors &  \texttt{(\myboldcode{defselector} $\sigma$ $\kappa$
  $[$ \myboldcode{:formals} $\Psi$ $]$ :$f_1$ $\epsilon_1$ \textrm{...} :$f_n$ $\epsilon_n$)} 
& define an selector $\iota$ of class $\kappa$ (usually
\mymelt{class\_selector}) with each extra field $f_i$
initialized to the value of $\epsilon_i$ {\relsize{-0}{(usually no extra fields are
given so $n=0$)}} and with optional formals $\Psi$
\\
\hline
for primitives &  \texttt{(\myboldcode{defprimitive} $\nu$ $\phi$ :$\theta$
  $\eta$)} & define primitive $\nu$ \petitun{with
formal arguments $\phi$, result c-type $\theta$ by macro-string
expansion $\eta$} \\
\hline
for c-iterators & \texttt{(\myboldcode{defciterator} $\nu$ $\Phi$
  $\sigma$ $\Psi$
  $\eta$ $\eta'$)} & define c-iterator $\nu$ \petitdemi{with
  input formals $\Phi$, state symbol $\sigma$,  local
  formals $\Psi$, start expansion $\eta$, end
  expansion $\eta'$} \\
\hline
for c-matchers & \texttt{(\myboldcode{defcmatcher} $\nu$ $\Phi$ $\Psi$
  $\sigma$ \petitdemi{$\eta$ $\eta'$})} & define c-matcher $\nu$ \petitdemi{with
  input formals $\Phi$ \emph{[the matched thing, then other inputs]}, output
  formals $\Psi$, state symbol $\sigma$, test expansion $\eta$, fill
  expansion $\eta'$} \\
\hline
for fun-matchers &  \texttt{(\myboldcode{defunmatcher} $\nu$ $\Phi$
  $\Psi$ $\epsilon$)} & define funmatcher $\nu$ \petitdemi{with
  input formals $\Phi$, output
  formals $\Psi$, with function $\epsilon$} \\
\hline
\hline
\multicolumn{3}{c}{\textbf{expressions exporting names}} \\
of values & \texttt{(\myboldcode{export\_value} $\nu_1$ \textrm{...})} &
export the names $\nu_i$ as bindings of values \petitdemi{(e.g., of
  functions, objects, matcher, selector, ...)}\\
\hline
of macros &  \texttt{(\myboldcode{export\_macro} $\nu$ $\epsilon$)} &
export name $\nu$ as a binding of a macro \petitdemi{(expanded by the $\epsilon$ function)} \\
\hline
of classes &  \texttt{(\myboldcode{export\_class} $\nu_1$ \textrm{...})}
& export every class name $\nu_i$ and all their own fields \petitun{(as value
bindings)} \\
\hline
as synonym &  \texttt{(\myboldcode{export\_synonym} $\nu$ $\nu'$)} &
export the new name $\nu$ as a synonym of the existing name $\nu'$ \\
\end{tabular}
\end{small}

Macro-expansion is internally the first step of \melt translation to
C: parsed (or in-heap) S-exprs (of \mymelt{class\_sexpr}) are
macro-expanded into a \melt ``abstract syntax tree'' (a subclass of
\mymelt{class\_source}). This macro machinery is extensively used,
e.g., \myboldcode{let} and \myboldcode{if} constructs are
macro-expanded (to instances of \mymelt{class\_source\_let} or
\mymelt{class\_source\_if} respectively.

Field names and class names are supposed to be globally unique, to
enable checking their access or update. Conventionally class names
start with \mymelt{class\_} and field names usually share a common
unique prefix in their class. There is no protection (i.e. visibility
restriction like \myboldcode{private} in C++) for accessing a field.

\medskip

All definitions accept documentation annotation using
\myboldcode{:doc}, and a documentation generator mode produces
documentation with-cross references in \prodname{Texinfo} format.

\bigskip
Miscellanous constructs are available, to help in debugging or coding or
to generate various C code depending on compile-time conditions.

\begin{small}
\relsize{-0.7}
\begin{tabular}{l|l|p{0.56\textwidth}}
\multicolumn{3}{c}{\textbf{expressions for debugging}} \\
debug message & \texttt{(\myboldcode{debug\_msg} $\epsilon$ $\mu$)} &
debug printing  message $\mu$ \& value $\epsilon$ \\
\hline
assert check &  \texttt{(\myboldcode{assert\_msg} $\mu$ $\tau$)} & nice
``halt'' showing message $\mu$ when asserted test $\tau$ is false \\
\hline
warning & \texttt{(\myboldcode{compile\_warning} $\mu$ $\epsilon$)} &
like \texttt{\#warning} in C: emit warning $\mu$ at \melt translation
time and gives $\epsilon$ \\
\hline
\multicolumn{3}{c}{\textbf{meta-conditionals}} \\
Cpp test &  \texttt{(\myboldcode{cppif} $\sigma$ $\epsilon$
  $\epsilon'$)} & conditional on a preprocessor symbol:
\petitun{emitted C code is \texttt{\#if $\sigma$} \fbox{code for
$\epsilon$} \texttt{\#else} \fbox{code for $\epsilon'$} \texttt{\#endif}} \\
\hline 
Version test &   \texttt{(\myboldcode{gccif} 
    $\beta$ $\epsilon_1$ \textrm{...})} & the $\epsilon_i$ are
  translated only if \gcc has version prefix
  string $\beta$ \\
\end{tabular} 
\end{small}

\bigskip
Reflective access to the current and parent environment is possible
(but useful in exceptional cases, since \textit{\texttt{export\_}...}
directives are available to extend the current exported environment):

\begin{small}
\begin{tabular}{l|l|p{0.3\textwidth}}
\multicolumn{3}{c}{\textbf{introspective expressions}} \\
Parent environment & \texttt{(\myboldcode{parent\_module\_environment})}
& \petitun{gives the previous module environment} \\
\hline
Current environment & \texttt{(\myboldcode{current\_module\_environment\_container})}
& \petitun{gives the container of the current module's environment} \\
\end{tabular}
\end{small}

\subsection{Linguistic constructs to fit MELT into GCC}
\label{subsec:lingconstruct}

Several language constructs are available to help fit \melt into
\gcc, taking advantage of \melt and \gcc runtime infrastructure
(notably \ggc). They usually use macro-strings to provide C code with
holes. Code chunks (\S \ref{subsub:code-chunk}) simply permit to insert C code in \melt
code. Higher-level constructs describe how to translate other \melt
expressions into C: primitives (\S \ref{subsub:primitives}) describe how to translate low-level
operations into C; c-iterators (\S \ref{subsub:c-iterators}) define how iterative expressions are
translated into \texttt{for}-like loops; c-matchers (\S \ref{sub:c-matchers}) define how to
generate simple patterns (for matching), etc.

\subsubsection{Code chunks} 
\label{subsub:code-chunk}

Code chunks are simple \melt templates (of \myboldcode{:void} c-type) for
generated \emph{C} code. They are the lowest possible way of impacting \melt
\emph{C} code generation, so are seldom used in \melt (like \texttt{\bf asm} is
rarely used in \emph{C}).

As a trivial example where \texttt{i} is a \melt
\myboldcode{:long} variable bound in an enclosing \myboldcode{let},
\begin{myalltt}(\textbf{code\_chunk} sta
  \#\{\$sta\#\_lab: printf("i=\%ld\textbackslash{n}", \$i++); goto \$sta\#\_lab; \}\# )
\end{myalltt}%
would be translated to 
\begin{myalltt}
\{sta\_1\_lab: printf("i=\%ld\textbackslash{n}", curfnum[3]++); goto sta\_1\_lab;\}
\end{myalltt}
the first time it translated (\mytexttt{i} becoming
\mytexttt{curfnum[3]} in C), but would use \mytexttt{sta\_2\_lab} the
second time, etc. 
The first argument of \mytexttt{\textbf{code\_chunk}}
- \mytexttt{sta} here - is a \emph{state} symbol, expanded to a C
identifier unique to the code chunk's translation. The second argument
is the macro-string serving as template to the generated C
code. The state symbol is uniquely expanded, and other symbols should
be \melt variables and are replaced by their translation.
So the \mytexttt{code\_chunk} of state symbol \mytexttt{helloworldchunk} in \S \ref{sub:macro-strings} is
translated into the following \emph{C} code:
\begin{myalltt}
  int i=0; /* our HELLOWORLDCHUNK__1 */ 
     HELLOWORLDCHUNK__1_label: printf("hello world from MELT\textbackslash{n}");
     if (i++ < 3) goto HELLOWORLDCHUNK__1_label; ;
\end{myalltt}

\subsubsection{Primitives} 
\label{subsub:primitives}

Primitives define a \melt operator by its C expansion. The unary
negation \mytexttt{negi} is defined exactly as :%
\begin{myalltt}
(\textbf{defprimitive} negi (\textbf{:long} i)  \textbf{:long} 
  \textbf{:doc} #\{\textit{Integer unary negation of \$i.}\}#
  #\{(-(\$i))\}#  )
\end{myalltt}

Here we specify that the formal argument \mytexttt{i} is, like the
result of \mytexttt{negi}, a \myboldcode{:long} stuff. We give an
optional documentation, followed by the macro-string for the C
expansion. Primitives don't have state variables but are subject to
normalization\footnote{Assuming that \mytexttt{x} is a \melt variable
  for a \myboldcode{:long} stuff, then the expression \mytexttt{(+i
    (negi x) 1)} is normalized as \mytexttt{\textbf{let} $\alpha = -x,
    \beta = \alpha + 1$ \textbf{in} $\beta$} in pseudo-code - suitably
  represented inside \melt (where $\alpha, \beta$ are fresh gensym-ed
  variables).} and type checking. During expansion, the formals
appearing in the primitive definition are replaced appropriately.

\subsubsection{C-iterators}
\label{subsub:c-iterators}

A \melt \emph{c-iterator} is an operator translated into a
\mytexttt{for}-like C loop. The \gcc compiler defines many constructs
similar to C \mytexttt{for} loops, usually with a mixture of macros
and/or trivial inlined functions. C-iterators are needed in \melt
because the \gcc API defines many iterative conventions.  For example,
to iterate on every \mytexttt{gimple} \textit{g} inside a given
\mytexttt{gimple\_seq} \textit{s} \gcc mandates (see \S
\ref{subsec:gcc-legacy}) the use of a
\mytexttt{gimple\_simple\_iterator}.

In \melt, to iterate on the \mytexttt{\textbf{:gimpleseq} s} obtained by
the expression $\sigma$ and do something on every
\mytexttt{\textbf{:gimple} g} inside \mytexttt{s}, we can simply code
\mytexttt{(\textbf{let} ( (\textbf{:gimpleseq} s $\sigma$) )
  (each\_in\_gimpleseq (s) (\textbf{:gimple} g) \quasicode{do
    something with \texttt{g}...}))} by invoking the \emph{c-iterator}
\mytexttt{each\_in\_gimpleseq}, with a list of inputs - here simply
\mytexttt{(s)} - and a list of local formals - here
\mytexttt{(\textbf{:gimple} g)} - as the iterated \emph{thing}s.

\smallskip
This c-iterator (a template for such \mytexttt{\bf for}-like loops) is defined exactly as:%
\begin{myalltt}
(\textbf{defciterator} each_in_gimpleseq
  (\textbf{:gimpleseq} gseq)                     \textit{;start formals}
  eachgimplseq                          \textit{;state}
  (\textbf{:gimple} g)                           \textit{;local formals}
  #\{\textit{/* start \$eachgimplseq: */}
   gimple_stmt_iterator gsi_$eachgimplseq;
   if ($gseq) for (gsi_$eachgimplseq = gsi_start ($gseq);
                   !gsi_end_p (gsi_$eachgimplseq);  
                   gsi_next (&gsi_$eachgimplseq)) \{
    $g  = gsi_stmt (gsi_$eachgimplseq);   \}#
  #\{ \} \textit{/* end \$eachgimplseq*/ } \}#)
\end{myalltt}
We give the start formals, state symbol, local formals and the
``before'' and ``after'' expansion of the generated loop block. The
expansion of the body of the invocation goes between the before and
after expansions. C-iterator occurrences are also normalized (like
primitive occurrences are). \melt expressions using c-iterators give a
\myboldcode{:void} result, since they are used only for their side
effects.

\subsection{Modules, environments, standard library and hooks}
\label{sub:module-env}

A single \texttt{*.melt} source file\footnote{\melt can also translate
  into C a sequence of S-expressions from memory, and then
  dynamically load the corresponding temporary module after it has
  been C-compiled.} is translated into a single module
loaded by the \melt run-time. The module's generated
\mytexttt{start\_module\_melt} routine [often quite big] takes
a parent environment, executes the top-level forms, and finally
returns the newly created module's environment. Environments and their
bindings are reified as objects.

\begin{sloppypar}
Only exported names add bindings in the module's environment. \melt
code can explicitly export defined values (like instances, selectors,
functions, c-matchers, \ldots) using the
\mytexttt{(\textbf{export\_values} \textrm{...})} construct; macros
(or pat-macros [that is pattern-macros producing abstract syntax of
  patterns]) definitions are exported using the
\mytexttt{(\textbf{export\_macro} \textrm{...})} construct or
\mytexttt{(\textbf{export\_patmacro} \textrm{...})}; classes and their
own fields are exported using the \mytexttt{(\textbf{export\_class}
  \textrm{...})}  construct. Macros and pattern macros in \melt are
expanded into an abstract syntax tree (made of objects of sub-classes
of \mymelt{class\_source}, e.g., instances of
\mymelt{class\_source\_let} or of \mymelt{class\_source\_apply},
\ldots), not into s-expressions (i.e. objects of \mymelt{class\_sexpr}, as
provided by the reader).
\end{sloppypar}

Field names should be \emph{globally} unique: this enables
\mytexttt{(\textbf{get\_field} :named\_name x)} to be safely translated
into something like ``if \mytexttt{x} is an instance of
\mymelt{class\_named} fetch its \mytexttt{:named\_name} field otherwise
give nil'', since \melt knows that \mytexttt{named\_name} is a field of
\mymelt{class\_named}.

As in C, there is only one name-space in \melt which is technically,
like Scheme, a Lisp$_1$ dialect\footnote{Each bound name is bound only
  once, and there are no separate namespaces like in C or Common
  Lisp.} (in Queinnec's terminology \cite{queinnec-lisp-1996}). This
prompts a few naming conventions: most exported names of a module
share a common prefix; most field names of a given class share the
same prefix unique to the class, etc.

The entire \melt translation process \cite{starynkevitch-2009-grow} is
implemented through many exported definitions which can be used by
expert \melt users to customize the \melt language to suit their
needs. Language constructs \footnote{Like
  \mytexttt{(\textbf{current\_module\_environment\_container})} and
  \mytexttt{(\textbf{parent\_module\_environment})}, etc.} give total
access to environments (instances of \mymelt{class\_environment}).

\sloppypar Hooks for changing \gcc's behavior are provided
on top of the existing \gcc plugin hooks (for instance, as exported
primitives like \mytexttt{install\_melt\_gcc\_pass} which installs
a \melt instance describing a \gcc pass and registers it inside \gcc).

A fairly extensive \melt standard library is available (and is used by
the \melt translator), providing many common facilities (map-reduce
operations; debug output methods; run-time asserts printing the \melt
call stack on failure; translate-time conditionals emitted as
\mytexttt{\#ifdef}; \ldots) and
interfaces to \gcc internals. Its \mytexttt{.texi} documentation is
produced by a generator inside the \melt translator.

When \gcc will provide additional hooks for plugins, making them
available to \melt code should hopefully be quite easy.


\section{Pattern matching in \MELT}
\label{sec:pattern}

Pattern matching
\cite{lefessant-maranget-2001,warn-jfp-07,pitrat-oneself-1995,wadler-1987-patternviews}
is an essential operation in symbolic processing and formal handling
of programs, and is one of the buying features of high-level
programming languages (notably Ocaml and Haskell). Several tasks
inside \gcc are mostly pattern matching (like simplification and
folding of constant expressions)\footnote{Strangely, \gcc has several
  specialized code generators, but none for pattern matching: so the
  file \mytexttt{gcc/fold-const.c} is hand-written (16\kloc).}. Code
using \melt pattern matching facilities is much more concise than its
(generated or even hand-written) \emph{C} equivalent.

\subsection{Using patterns in \MELT}
\label{sub:using-patterns}

Developers using \melt often need to filter complex \gcc \emph{stuff} (in
particular \texttt{gimple} or \texttt{tree}-s) in their \gcc passes
coded in \melt. This is best achieved with pattern matching. The
matching may fail (if the data failed to pass the filter) or may
extract information from the matched data.

\subsubsection{About pattern matching}

Patterns are major syntactic constructs (like expressions and
let-bindings in Scheme or \melt). In \melt, a pattern starts with a
question mark, which is parsed particularly: \mytexttt{?x} is the same
as \mytexttt{(\textbf{question} x)} [it is the pattern variable
  \textit{x}]. \texttt{?\_} is \footnote{\texttt{?\_} can be pronounced
  as ``joker''} the \emph{wildcard pattern} (matching anything). An
expression occurring in pattern context is a \emph{constant}
pattern. Patterns may be nested (in composite patterns) and occur in
\myboldcode{match} expressions.

Elementary patterns are ultimately translated into code that
\emph{tests} that the matched \emph{thing} $\mu$ can be filtered by the
pattern $\pi$ followed by code which extracts appropriate data from
$\mu$ and \emph{fills} some locals with information extracted from
$\mu$. Composite patterns need to be translated and optimized to
avoid, when possible, repetitive tests or fills.

\subsubsection{An example of pattern usage in \gccmelt}
\label{para:exemple-pattern}
\sloppy
Many tasks depend upon the form of [some intermediate internal
  representation of] user source code, and require extracting some of
its sub-components.  For instance, the author has written (in a single
day) a \gcc extension in \melt to check simple coding rules in
\mytexttt{melt-runtime.c}, (e.g., in function of figure \ref{fig:newboxint}). When enabled with
\mytexttt{-fplugin-melt-arg-mode=meltframe}, it adds a new pass (after the \texttt{"ssa"} pass\footnote{\texttt{ssa} means Static Single Assignment, so at that stage the code is represented in \emph{Gimple/SSA} form, so each SSA variable is assigned once!}. of \gcc \cite{pop-phd2006-ssa})
\mytexttt{melt\_frame\_pass} to \gcc. This pass first finds the
declaration of the local \mytexttt{meltfram\_\_} in the following
pass execute function:
\begin{quote}\begin{small}
\begin{relsize}{-1}
\begin{Verbatim}[numbers=left,frame=lines,commandchars=\\\{\}]
(defun meltframe_exec (pass)
  (let ( 
      (:tree tfundecl (cfun_decl))           (:long nbvarptr 0)
      (:tree tmeltframdecl (null_tree))      (:tree tmeltframtype (null_tree)) )
    (each_local_decl_cfun ()   (:tree tlocdecl :long ix)
      (match tlocdecl 
             ( ?(tree_var_decl
                 ?(and ?tvtyp ?(tree_record_type_with_fields ?tmeltframrecnam ?tmeltframfields))
                 ?(cstring_same "meltfram__") ?_)
               (setq tmeltframdecl tlocdecl)    (setq tmeltframtype tvtyp)
               (foreach_field_in_record_type (tmeltframfields) (:tree tcurfield)
                 (match tcurfield
                        ( ?(tree_field_decl
                            ?(tree_identifier ?(cstring_same "mcfr_varptr"))
                            ?(tree_array_type ?telemtype 
                                              ?(tree_integer_type_bounded ?tindextype
                                                                          ?(tree_integer_cst 0)
                                                                          ?(tree_integer_cst ?lmax) 
                                                                          ?tsize)))
                           (setq tmeltframvarptr tcurfield)  (setq nbvarptr lmax)))))
             ( ?_ (void))))
\end{Verbatim}
\end{relsize}

The \mytexttt{let} line 2 spans the entire \melt function
\mytexttt{meltframe\_exec}, with bindings lines 3 \& 4 for
\mytexttt{tfundecl}, \mytexttt{nbvarptr}, \mytexttt{tmeltframdecl} \&
\mytexttt{tmeltframtype} locals. The
\mytexttt{each\_local\_decl\_cfun} is a c-iterator (iterating -lines 5
to 11- on the \emph{Tree}-s representing the local declarations in the
function). The \mytexttt{match} expression filters the current local
declaration \texttt{tlocdecl} (lines 7-11). When it is a variable
declaration (line 7) whose type matches the sub-pattern line 8 and
whose name (line 9) is exactly \texttt{meltfram\_\_}, we assign (line
10) appropriately \mytexttt{tmeltframdecl} \&
\mytexttt{tmeltframtype}, and we iterate (line 11) on its fields to
find, by the \mytexttt{match} (lines 12-21), the declaration of field
\mytexttt{mcfr\_varptr} (in the \emph{C} code), and its array index
upper bound \mytexttt{lmax}, assigning them (line 20) to locals
\mytexttt{tmeltframvarptr} \& \mytexttt{nbvarptr}. Otherwise, using
the wildcard pattern \mytexttt{?\_}, we give a \mytexttt{:void} result
for the match of \mytexttt{tlocdecl} (line 21).
\end{small}
\end{quote}

\medskip
Once the declaration of \mytexttt{meltfram\_\_} and of its
\mytexttt{mcfr\_varptr} field has been found\footnote{A warning is issued if \mytexttt{meltfram\_\_} or  \mytexttt{mcfr\_varptr} has not been found.} in the current function
(given by \mytexttt{cfun} inside \gcc), we iterate on each basic block
\mytexttt{bb} of that function, and on each \emph{gimple} statement
\mytexttt{g} of that basic block, and we match that statement \mytexttt{g} to find assignments to or from \mytexttt{meltfram\_\_.mcfr\_varptr[$\kappa$]} where $\kappa$ is some constant integer index:

\begin{quote}\begin{small}
\begin{relsize}{-1}
\begin{Verbatim}[numbers=left,frame=lines,firstnumber=last,commandchars=\\\{\}]
      (each_bb_cfun  ()   (:basic_block bb  :tree fundecl)
       (eachgimple_in_basicblock (bb)
        (:gimple g)
        (match g
               ( ?(gimple_assign_single
                   ?(tree_array_ref  ?(tree_component_ref tmeltframdecl tmeltframvarptr)
                     ?(tree_integer_cst ?idst))
                   ?(tree_array_ref  ?(tree_component_ref  tmeltframdecl  tmeltframvarptr)
                     ?(tree_integer_cst ?isrc)))
                 \relsize{+0.5}{\quasicode{handle assign ``\mytexttt{meltfram__.mcfr_varptr[\textit{idst}] = meltfram__.mcfr_varptr[\textit{isrc}];}''}})
               ( ?(gimple_assign_single
                   ?(tree_array_ref  ?(tree_component_ref tmeltframdecl tmeltframvarptr)
                     ?(tree_integer_cst ?idst))
                   ?rhs)
                 \relsize{+0.5}{\quasicode{handle assign ``\mytexttt{meltfram__.mcfr_varptr[\textit{idst}] = \textit{rhs};}''}})
               ( ?(gimple_assign_single ?lhs
                   ?(tree_array_ref  ?(tree_component_ref tmeltframdecl tmeltframvarptr)
                     ?(tree_integer_cst ?isrc)))
                 \relsize{+0.5}{\quasicode{handle assign ``\mytexttt{\textit{lhs} = meltfram__.mcfr_varptr[\textit{isrc}];}''}})
\end{Verbatim}
\end{relsize}

The gimple \mytexttt{g} is matched against the most filtering pattern (lines 26-30, for assignments like ``\mytexttt{meltfram\_\_.mcfr\_varptr[\textit{idst}] = meltfram\_\_.mcfr\_varptr[\textit{isrc}];}'' ) first, then against the more general patterns -for ``\mytexttt{meltfram\_\_.mcfr\_varptr[\textit{idst}] = \textit{rhs};}'' where \textit{\mytexttt{rhs}} is any simple operand- lines 32-36, and for ``\mytexttt{\textit{lhs} = meltfram\_\_.mcfr\_varptr[\textit{isrc}];}'' lines 37-40. The \melt programmer should order his matching clauses from the more specific to the more general.
  \end{small}
\end{quote}

Other code (not shown here) in function \mytexttt{meltframe\_exec} remembers all
left-hand side and right-hand side occurences of
\mytexttt{meltfram\_\_.mcfr\_varptr[$\kappa$]}, and issues a warning when such a slot is not used.

\bigskip
We see that a \texttt{\textbf{match}} is made of several match-cases,
tested in sequence until a match is found. Each case starts with a
pattern, followed by sub-expressions which are computed with the
pattern variables of the case set appropriately by the matching of the
pattern; the last such sub-expression is the result of the entire
\texttt{\textbf{match}}. Like other conditional forms in \melt,
\myboldcode{match} expressions can give any \emph{thing} (\emph{stuff}, e.g.,
\myboldcode{:long} \ldots or even \myboldcode{:void}, or \emph{value}) as
their result.  Patterns may be nested like the
\mytexttt{tree\_var\_decl} or \mytexttt{tree\_record\_type} above.
All the locals for pattern variables in a given match-case are cleared
(before testing the pattern). It is good style to end a
\mytexttt{\textbf{match}} with a catch-all wildcard \mytexttt{?\_}
pattern.

A pattern is usually composite (with nested sub-patterns) and has a
double role: first, it should \emph{test} if the matched \emph{thing} fits;
second, when it does, it should extract \emph{thing}s and transmit them to eventual
sub-patterns; this is the \emph{fill} of the pattern. The matching of
a pattern should conventionally be without side-effects (other than
the fill, i.e. the assignment of pattern variables).

Patterns may be \emph{non-linear}: in a matching case, the same
pattern variable can occur more than once; then it is set at its first
occurrence, and tested for \emph{identity}\footnote{We don't test for
  \emph{equality} of values or other \emph{thing}s, knowing that
  $\lambda$-term equality is undecidable, and acknowledging that deep
  equality compare of ASTs like \mytexttt{tree} or \mytexttt{gimple}
  is too expensive.} with \mytexttt{==} in the generated \emph{C} code on all
the following occurrences. This is useful in patterns like
\mytexttt{?(gimple\_assign\_single ?var ?var)} to find assignments of
a variable \texttt{var} to itself.

\subsection{Pattern syntax overview}

A pattern $\pi$ may match some matched \emph{thing} $\mu$, or may fail. It
the matching succeeds, sub-patterns may be matched, and pattern
variables may become bound. The \emph{thing} bound by some pattern variable
is checked in following occurrences of the same pattern variables and
is available inside the match-clause body.

Patterns may be one of:

\begin{itemize}

\item expressions $\epsilon$ \petitun{(e.g., constant literals)} are (degenerated)
patterns. They match the matched data $\mu$ iff $\epsilon$ \texttt{==}
$\mu$ (for the C sense of equality, which for pointers is their identity).

\item The \textbf{wildcard} noted {\large \meltcodec{?\_}} matches
  everything (every \emph{value} or \emph{stuff}) and never fails.

\item a pattern variable \meltcodec{?}$\nu$ matches $\mu$ if it was
  unset \petitdemi{(by a previous [sub-]matching of the same
  \meltcodec{?}$\nu$)}. In addition, it is then bound to $\mu$.  If the
  pattern variable was previously set, it is tested for identity (with
  equality in the C sense).

\item most patterns are \textbf{matcher} patterns  \texttt{?($m$
  $\epsilon_1$ \textrm{...} $\epsilon_n$ $\pi_1$ \textrm{...} $\pi_p$)} where the $n \ge
  0$ expressions $\epsilon_i$ are input parameters to the matcher $m$
  and the $\pi_j$ sub-patterns are passed extracted data. The matcher
  is either a \emph{c-matcher} (declaring how to translate that pattern to
  \emph{C} code) or it is a \emph{fun-matcher} (matching is done by a \melt function returning secondary \emph{thing}s).

\item instance patterns are like \texttt{?(\myboldcode{instance} $\kappa$  \textbf{:$\Phi_1$}
    $\pi_1$ \textrm{...} \textbf{:$\Phi_n$}
    $\pi_n$)}; the matched $\mu$ is an object of [a sub-] class $\kappa$
  whose field $\Phi_i$ matches sub-pattern $\pi_i$.

\item conjunctive patterns are \texttt{?(\myboldcode{and} $\pi_1$
  \textrm{...} $\pi_n$)} and they match $\mu$ iff every $\pi_i$ in sequence
  matches $\mu$; notice that when some $\pi_i$ is a pattern variable \meltcodec{?}$\nu$ that variable is matched and $\mu$ should match the further  $\pi_j$ (with $j > i$) with $\nu$ appropriately bound to $\mu$. (This generalizes the \texttt{as} keyword inside Ocaml patterns).

\item disjunctive patterns are \texttt{?(\myboldcode{or} $\pi_1$
  \textrm{...} $\pi_n$)} and they match $\mu$ if one of the $\pi_i$ matches $\mu$.
\end{itemize}

\subsection{C-matchers and fun-matchers}
\label{sub:c-matchers}

The \emph{c-matchers} are one of the building blocks of patterns -
much like primitives are one of the building blocks of
expressions. Like primitives, c-matchers are defined as a specialized
C code generation template. In the example above (\S
\ref{para:exemple-pattern}), most composite patterns involve
c-matchers: \mytexttt{tree\_var\_decl}, \mytexttt{tree\_record\_type} and
\mytexttt{cstring\_same} are C-matchers.

Like for every pattern, a C-matcher defines how the pattern using it
should perform its test, and then how it should do its fill.
A simple example of a C-matcher is \mytexttt{cstring\_same}: some
\myboldcode{:cstring} \emph{stuff} $\sigma$ matches the pattern
\mytexttt{?(cstring\_same "fprintf")} iff $\sigma$ is the same as the
\mytexttt{const char*} string \mytexttt{"fprintf"} given as input to our
c-matcher. This c-matcher has a test part, but no fill part (because
used without sub-patterns).%
\begin{small}
\begin{myalltt}
(\textbf{defcmatcher} cstring_same  (\textbf{:cstring} str cstr) () strsam
  \textbf{:doc} #\{\textit{The \$CSTRING\_SAME c-matcher matches a string \$STR iff it equals the constant string \$CSTR. 
         The match fails if \$STR is null or different from \$CSTR.}\}# 
  #\{ /*\$STRSAM test*/ (\$STR != (const char*)0 && \$CSTR != (const char*)0 && !strcmp(\$STR, \$CSTR)) \}# )
\end{myalltt}
\end{small}
Notice that the state symbol \mytexttt{strsam} is used inside a
comment, to uniquely identify each occurrence in the generated C, and
that we take care of testing against null \texttt{const char*}
pointers to avoid crashes.

\smallskip
A more complex (and \gcc specific) example is the
\mytexttt{gimple\_assign\_single} c-matcher (to filter single assignments
in compiled code). It defines both a testing and a filling expansion
using two macro-strings:%
\begin{myalltt}
(\textbf{defcmatcher} gimple_assign_single
   (\textbf{:gimple} ga)  (\textbf{:tree} lhs rhs)  gimpassi
   #\{ /*\$GIMPASSI test*/(\$GA && gimple_assign_single_p (\$GA)) \}#
   #\{ /*\$GIMPASSI fill*/ \$LHS = gimple_assign_lhs (\$GA);   \$RHS = gimple_assign_rhs1(\$GA); \}# )
\end{myalltt}
Here \mytexttt{ga} is the matched gimple, and \mytexttt{lhs} \& \mytexttt{rhs} are the
output formals: they are assigned in the fill expansion to transmit
\mytexttt{tree}-s to sub-patterns!

C-matchers are a bit like Wadler's notion of \emph{Views}
\cite{wadler-1987-patternviews}, but are expanded into C code. \melt
also has \emph{fun-matchers} which similarly are views defined by a
\melt function returning a non-nil value if the test succeeded with
several secondary results giving the extracted \emph{thing}s to sub-patterns.
For example the following code defines a fun-matcher
\mytexttt{isbiggereven}\footnote{Our \mytexttt{isbiggereven} could
  also be defined as a c-matcher!} such that the pattern
\mytexttt{?(isbiggereven $\mu$ $\pi$)} matches a \myboldcode{:long}
stuff $\sigma$ iff $\sigma$ is a even number, greater than the number
$\mu$, and $\sigma / 2$ matches the sub-pattern $\pi$. We define an
auxiliary function \mytexttt{matchbiggereven} to do the matching [we
  could have used a \myboldcode{lambda}]. If the match succeeds, it
returns a true (i.e. non nil) value (here \mytexttt{fmat}) and the
integer to be matched with $\pi$. Its first actual argument is the
fun-matcher \mytexttt{isbiggereven} itself. The testing behavior of
the matching function is its first result (nil or not), and the fill
behavior is through the secondary results.

\begin{myalltt}
(\textbf{defun} matchbiggereven (fmat \textbf{:long} s m) 
\textit{; fmat is the funmatcher, s is the matched \(\sigma\), m is the minimal \(\mu\)} 
   (\textbf{if} (==i (\%iraw s 2) 0) 
      (\textbf{if} (>i s m) (\textbf{return} fmat (/iraw m 2)))))
(\textbf{defunmatcher} isbiggereven (\textbf{:long} s m) (\textbf{:long} o) matchbiggereven)
\end{myalltt}

The fun-matcher definition has an input formals list and an output
formal list, together defining the expected usage of the fun-matcher
operator in patterns.

Both c-matchers and fun-matchers can also define what they mean in
expression context (not in pattern one). So the same name can be used
for constructing expressions and for destructuring patterns.

\subsection{Implementing patterns in \MELT}
\label{sub:implementing-patterns}

Designing and implementing patterns in \melt was quite difficult,
because a good translation of pattern matching should :

\begin{itemize}
\item factorize, when possible, common sub-patterns, to avoid testing
  twice the same \emph{thing}.
\item share, when appropriate, data extracted from subpatterns.
\item preferably re-use the many temporary locals used by the
  translation of the match, to lower the current \melt stack frame
  size.
\end{itemize}

Our first implementation of pattern translation to C is quite naive, and
uses simple memoization techniques to factorize sub-patterns or share
extracted data.

\bigskip

\begin{figure}
\begin{center}
\includegraphics[width=0.82\textwidth]{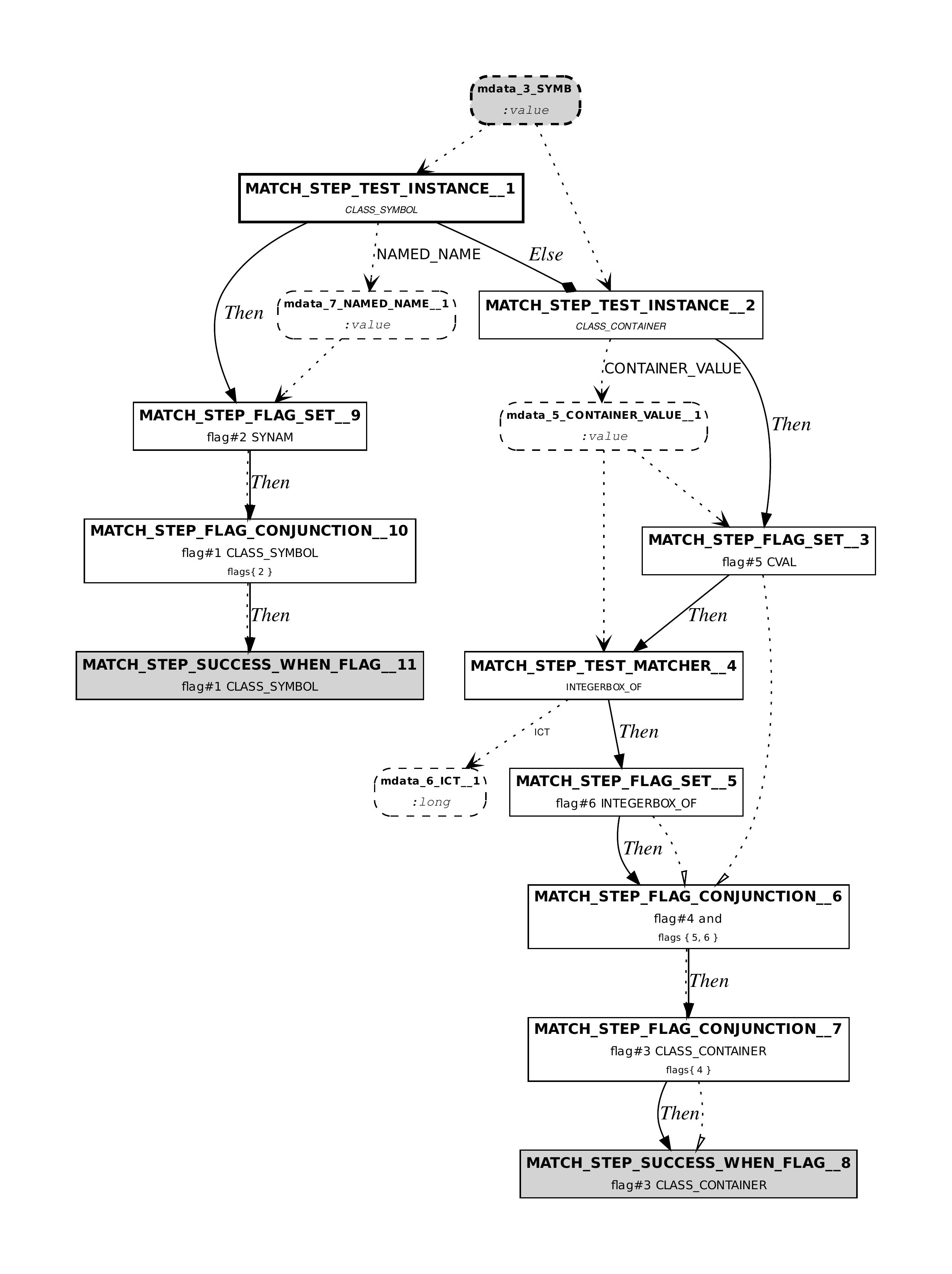}
\end{center}

\smallskip
\hrule

\raisebox{9ex}{\large Legend:} \hspace{3em} \includegraphics[height=0.13\textheight]{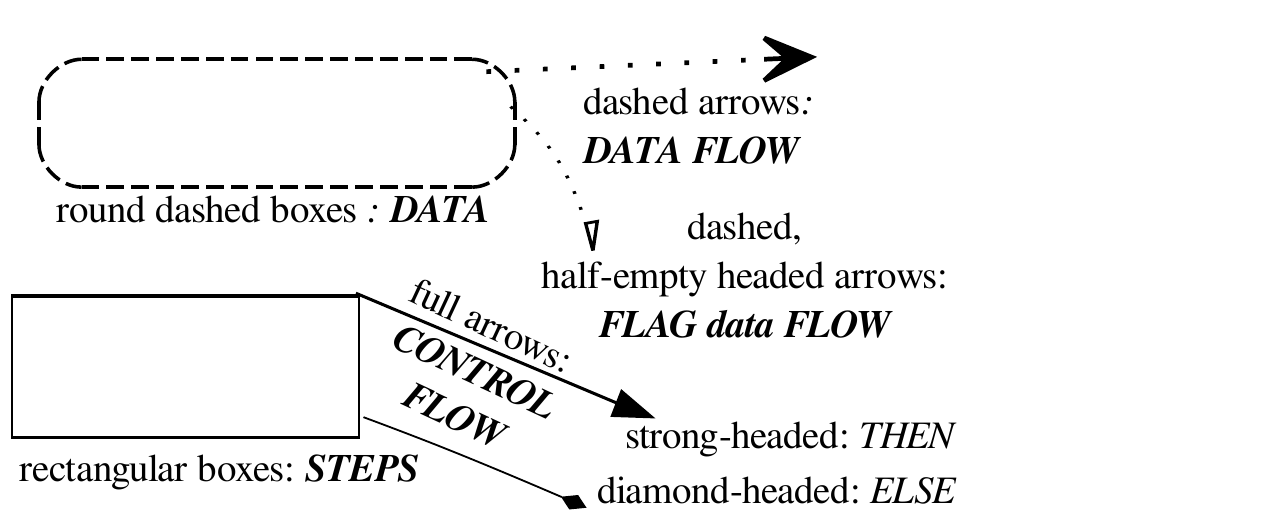}

\caption{\label{fig:pattern} internal graph for match}
\end{figure}

A better implementation of the pattern translator builds explicitly a
directed graph (with shared nodes for tests and data), like figure
\ref{fig:pattern}. The graph has data nodes (for temporary variables
for [sub-]matched \emph{thing}s, or for boolean flags internal to
the match) and elementary control steps. These steps are either tests
(with both a ``then'' and an ``else'' jumps to other steps) or
computations (usually with a single jump to a successor step). Some
steps just set an internal boolean flag, or compute the conjunction of
other flags. Other steps represent the testing or the filling parts of
c-matchers or fun-matchers. Final success steps correspond to
sub-expressions in the body of the matched clause and are executed if
a flag is set.

For instance a simple match (where \texttt{v} is the matched value) like below
is translated into the complex internal graph \footnote{To debug the
  pattern-match translator, \melt is generating a graph to be
  displayed with \prodname{GraphViz}. We have edited it (by removing
  details like source code location) for clarity.} given in figure
\ref{fig:pattern}:

\begin{small}
\begin{relsize}{-0.5}
\begin{alltt}
  (match v 
         ( ?(instance class_symbol :named_name  ?synam)
            (f synam))
         ( ?(instance class_container :container_value ?(and ?cval  ?(integerbox_of ?_)))
            (g cval)))
\end{alltt}
\end{relsize}
\end{small}
A more complex match like \mytexttt{(match
  tcurfield \textrm{...})} of \S \ref{para:exemple-pattern} code line
12-20 produces about 20 match steps and 12 match data.
This enhanced pattern matching is not entirely implemented at time of
writing: the generation of the control graph for the match is
implemented, but its translation into C is incomplete.


\section{Conclusions and future work}

Enhancing a legacy huge software with a domain specific language or
scripting language is always a major challenge (\S \ref{sec:intro}),
since incorporating a DSL inside a software is a major architectural
design decision which should be taken early. Mature big software like
\gcc have their coding habits, memory management strategies and data
organization which makes it very difficult to embed an existing
scripting language (like Python, Ocaml, Ruby, ...).

We have shown that adding a high-level DSL to a big software like \gcc
is still possible, by designing a run-time system
\S\ref{sec:using-runtime} compatible with the existing infrastructure
(notably \ggc) and most importantly, by having the DSL deal both with
boxed \emph{values} and raw existing \emph{stuff} in
\S\ref{subsec:values}. Translating the DSL to the language (with its
habits) used in that big software (\emph{C} for \gcc) enables
high-level language constructs in our DSL.  We have described a set of
language constructs in \S\ref{subsec:lingconstruct} (c-matchers,
primitives, c-iterators, \ldots) which give templates for \emph{C} code
generation.

Our empirical approach of designing and implementing a DSL like \melt
to fit into a large software like \gcc, could probably be re-used
for adding DSLs inside other huge mature software projects: designing
a runtime suitable for such a project, having several sorts of \emph{thing}s
(\emph{value}s and \emph{stuff}), generating code in the style of the existing
legacy, and defining adequate language constructs giving
code-generating templates.

\bigskip

Future work within \melt is mostly using this DSL to build interesting
\gcc extensions. P. Vittet has started in May 2011 a \emph{Google
  Summer of Code} project to add specific warnings into \gcc using
\melt. A. Lissy considers using it for Linux kernel
\cite{lissy-fosdem2011-modelchecking} code analysis.
The \texttt{opengpu} mode should be completed. Also, some language
features can be added or improved:
\begin{enumerate}
\item variadic functions, possibly provided by a \myboldcode{:rest}
  keyword similar to Common Lisp's \mytexttt{\&rest}. These should be
  very useful for debugging and tracing messages.
\item adding backtracking or iterating pattern constructs; for
  instance to be able to have a pattern for any
  \mytexttt{:gimple\_seq} \emph{stuff} containing at least one gimple
  matching a given sub-pattern.
\item adding a nice usable and hygenic macro system, inspired by
  Scheme's \texttt{defsyntax}
\item performance improvements might be achieved by sometimes
  translating \melt function calls into a C function call whose
  signature mimicks the \melt function signature.  
\item a message caching machinery, where every \melt message passing
  occurrence would use a cache (keeping the last class of the
  sending).
\item a central monitor, which would communicate with parallel
  \gccmelt compilations through asynchronous textual protocols.
\end{enumerate}

More generally, making \melt more high-level and more declarative (in J.Pitrat's \cite{pitrat-96-reflective,pitrat-beings-2009} sense) to
be able to express \gcc passes easily and concisely is an interesting
challenge, and could be transposed to other legacy software.

\section*{Acknowledgments}

Work on \melt has been funded by DGCIS thru ITEA GlobalGCC and FUI
OpenGPU projects.

Thanks to Albert Cohen, Jan Midtgaard, Nic Volanschi and to the
anonymous reviewers for their constructive suggestions and
their proof-reading. Residual mistakes are mine.


\nocite{*}
\bibliographystyle{eptcs}
\bibliography{MELT-Starynkevitch-DSL2011}

\end{document}